\documentstyle[12pt,aasms4]{article}

\lefthead{}
\righthead{}

\def\fasec{{\rlap.}^{\prime\prime} \hskip0.05em}

\def\amin{^{\prime}}
\def\asec{^{\prime\prime}}
\def\smpyr{\mbox{ M}_{\odot} \mbox{ yr}^{-1}}
\def\mdot{\stackrel{.}{\mbox{M}} }
\def\simlt{\; \stackrel{<}{_{\,\sim}}}

\begin{document}

\title{A Search for Radio Emission from Supernovae \\
With Ages from About One Week to More Than 80 Years}

\author{Christopher R. Eck,
John J. Cowan,
and David Branch}

\affil{Department of Physics and Astronomy,
University of Oklahoma, Norman, OK 73019}

\begin {center}
Christopher\_R\_Eck@raytheon.com, cowan@mail.nhn.ou.edu,
branch@mail.nhn.ou.edu
\end {center}

\begin{abstract}
We report VLA radio observations of 29 SNe with ages ranging from 10 days
to about 90 years past explosion.  These observations significantly
contribute to the existing data pool on such objects.  Included are
detections of known radio SNe~1950B, 1957D, 1970G, 1983N, the suspected
radio SN~1923A, and the possible radio SN~1961V.
None of the remaining 23 observations resulted in detections,
providing further evidence to support the observed trend that most SNe
are not detectable
radio emitters.  To investigate the apparent lack of radio emission from the
SNe reported here, we have followed standard practice and used Chevalier's
``standard model'' to derive (upper limits to) the mass-loss rates for the supernova
progenitors.  These upper limits
to the fluxes are consistent with a lack of circumstellar material
needed to provide detectable radio emission for SNe at these ages and distances.
Comparison of the radio luminosities of these supernovae as a function of
age past explosion to other well-observed radio SNe indicates that
the Type~II SNe upper limits are more consistent with the extrapolated
light curves of SN~1980K than of SN~1979C, suggesting that SN~1980K
may be a more typical radio emitter than SN~1979C.
For completeness, we have included an appendix where the results of analyses of the
non-SN radio sources are presented.  Where possible, we make (tentative)
identifications of these sources using various methods.
\end{abstract}


\section{INTRODUCTION}
Supernovae have been observed visually and recorded for at least the
last 1000 years (\cite{cla82}) but only in this century have we
been able to
observe these bright, energetic events in a waveband other than the
optical.  The different supernovae (SNe) are classified
into two main Supernova (SN) Types, I and II, distinguished
by features in their optical spectra.  The presence or absence of
other optical features classify SNe into several
sub-classifications, SNe~Ia, Ib, Ic, and some other
proposed sub-classes (e.g., the SNe~IIn [\cite{sch90}]).  To
understand SNe explosions and the mechanisms for the emission of radiation,
it is helpful to have knowledge of the pre-SN progenitor system.
While a basic understanding of the pre-SN progenitor stars
has been achieved (e.g., Branch, Nomoto, \& Filippenko 1991, \cite{bra95}),
the specifics are still a matter of some debate (e.g., \cite{nom96}).  Radio
observations of SNe have been shown (e.g., \cite{wei89}) to give insight into the
pre-SN progenitor system.

The first discovery of radio emission, from SN~1970G in NGC~5457 (M101)
(\cite{got72}, \cite{all76}), introduced this new regime in
which to study SNe.  Modern detections of well-observed radio SNe include
the Type II SNe~1978K in NGC~1313 (\cite{ryd93}, \cite{sch99}),
1979C in NGC~4321 (\cite{wei86}, \cite{wei91}, \cite{mon00}),
1980K in NGC~6946 (\cite{wei86}, \cite{wei92a}, \cite{mon98}),
1981K in NGC~4258 (\cite{wei86}, \cite{van92}, \cite{lac99}),
1986J in NGC~891 (\cite{rup87}, Weiler, Panagia, \& Sramek 1990),
1988Z in MCG~+03-28-022 (\cite{van93b}, \cite{lac99}),
and 1993J in NGC~3031 (M81) (\cite{van94}); the
Type Ib SNe~1983N in NGC~5457 (Sramek, Panagia, \& Weiler 1984,
\cite{cow85}) and 1984L in NGC~991 (Panagia, Sramek, \& Weiler 1986);
and the Type Ic SNe~1990B in NGC~4568 (\cite{van93a}).
Most of these SNe can now be considered as ``intermediate-age'' supernovae,
originally defined by Cowan \& Branch (1985) to have
an age between 10 and 100 years old.  These decades-old supernovae
are observed in the radio much after a supernova fades in the optical, typically
within two years, and before the earliest turn on of an SNR, predicted to take
at least 100 years (see also Cowan, Roberts, \& Branch 1994 and
Stockdale et al. 2001a, b for further discussion).
Many of the detected ``intermediate-age'' SNe whose ages are greater than 30 years
include SNe~1970G (Cowan, Goss, \& Sramek 1991; Stockdale et al. 2001a),
1968D in NGC~6946 (\cite{hym95}),
1961V in NGC~1058 (\cite{bra85}, Cowan, Henry, \& Branch 1988, \cite{sto01b})
[but there is some discussion on whether this was a SN explosion or the
pulse of an LBV (e.g., \cite{fil95})], 1957D \& 1950B in NGC~5457 (M83)
(Cowan et al. 1994, \cite{cow85}, \cite{cow82})
and the probable radio SN~1923A (\cite{eck98}).

The investigation of radio emission from Supernova Remnants (SNRs) has been well studied
due in part to the the strength, nature and longevity of the emission.  (While a
review of radio emission from SNRs is beyond the scope of this paper, a review of the
emission from SNRs can be found in Weiler \& Sramek (1988) and
Matonick \& Fesen (1997)).
Some examples of well-known galactic SNRs include SNR~1006, Kepler's SNR, and
Tycho's SNR as well as many extra-galactic SNRs (Green 1984).
One particular radio SNR of interest is Cas~A (\cite{baa77}),
the remnant from an SN that exploded in approximately
1670.  At about 300 years old, Cas~A is significant to our study because it is
the youngest radio SNR with a known age.  It is therefore useful for comparison
with observations of many of our older SNe in probing the relationship between
radio SNe and SNRs.

Despite attempts to look for radio emission from all SN types, no Type~Ia SNe
have been detected (e.g., \cite{eck95}, \cite{sra93}, \cite{wei89}), and
most Type~II or Ib/c SNe (i.e., core-collapse SNe) have not
been detected as radio emitters despite several searches
(e.g., \cite{eck96}, \cite{sra93}, \cite{wei89}).
However, when radio emission has been detected the Chevalier model
(Chevalier [1982a,b], 1984) describes the radio emission and fits the data very well
and may offer an explanation for the lack of detectable radio emission
from every SN.  In this model the radio emission is related to
circumstellar matter.  For older SNe that may be entering the SNR
phase, theoretical models (\cite{gul73}) and
calculations (\cite{cow84}) describe the radio emission
as related to the amount of interstellar matter (among other parameters
such as energy released in the explosion)
instead of circumstellar matter, but
the emission mechanisms in each case are similar.  While the Gull
model predicts a minimum time of about 100 years for a SNR to brighten,
it is noteworthy that SN 1885A has not been detected in the radio (Crane
et al. 1992) at an age of about 107 years.

In this paper, we present results of a long term survey to detect radio
emission from SNe at ages from 7 days to almost 90 years old.  Our results
confirm the observed trend that most SNe at these ages do not show radio emission.
In \S 2 we present our observations, and in \S 3 discuss our results
as applied to the Chevalier model to derive (upper limits to) mass-loss
rates.  We then discuss our results in relation to
others' observations and work.  In \S 4, we follow with conclusions.
In Appendix A we present the results of our analysis of the non-SNe
sources in our maps, and attempt to identify the sources when possible.

\section{OBSERVATIONS}
This series of observations began in an attempt to detect
radio emission from intermediate-age SNe.  If detected, a monitoring
program could begin at the Very Large Array (VLA)\footnote{The VLA is
a facility of the National Radio Astronomy Observatory, which is operated
by Associated Universities, Inc., under contract to the National Science
Foundation.} to trace the evolution from a radio supernova (RSN) into an SNR.
Since there is very
little data on the transition of RSNe to SNRs, these observations would help
in understanding this phase of SN evolution.

All observations were taken at the VLA over a time
span ranging from 1981 to 1995, and the SNe observed vary in distance and in
ages at the time of observation.  Tables~1-3 summarize relevant
observational data for the SNe reported here.  Table~1 gives parameters
for each observation, and Table~2 shows distances assumed for each SN, most
based on Cepheid observations in the parent galaxy or in a galaxy in the same
group.
Table~3 summarizes data on each SN, its parent galaxy, the age at the time of
observation and the flux densities for detections or $3\sigma$
upper limits to detection.  Table~4 presents the results of our analysis of the observational data (see \S 3.1 for discussion).

In addition to analyzing the region near the SNe sites, for completeness we
analyzed the non-SNe sources to determine a probable identity based on
superpositions with maps made at optical (H$\alpha$) wavelengths or on
spectral indices, for example.  Descriptions of the analyses of each map were
appended to the paper as Appendix~A to preserve the main focus of the paper
on radio SNe.
Table~5 lists the positions and fluxes for these non-SNe sources and
a possible identity for each.

\section{DISCUSSION}

\subsection{The Model}

To gain insights into the pre-SN progenitor system via radio emission, we begin
with the expression for the radio luminosity for Type~II SNe based on the
well-established Chevalier (1982a,b) mini-shell model.  According to this model,
the outward moving shock from the SN explosion encounters circumstellar
material, compresses it and creates Rayleigh-Taylor instabilities.
The magnetic fields locked in the material
compress and amplify, accelerating free electrons.  Synchrotron
radiation results and is responsible for the luminosity we see.
This luminosity has the form (following Chevalier [1982b])

\begin{equation}
L \propto \left(\frac{\stackrel{\cdot}{M}}{w}\right)^{(\gamma -7+12m)/4}
          t^{-(\gamma +5-6m)/2} m^{(5+\gamma)/2} U^{3(1-m)}
          \nu^{-(\gamma -1)/2} e^{-\tau_{\nu,ff}}
\end{equation}
where
\begin{equation}
          \tau_{\nu,ff} = C \left(\frac{\stackrel{\cdot}{M}}{w}\right)
                          ^{(5-3m)} t^{-3m} \nu^{-2.1} ,
\end{equation}
$\stackrel{\cdot}{\mbox{M}}$ is the mass-loss rate of the pre-SN progenitor
in $\smpyr$, $w$ is the wind speed in km s$^{-1}$, $\gamma$ is the electron
energy index, related to the radio spectral index, $\alpha$, by $\gamma =
2\alpha +1$, $t$ is the time since explosion in days, $m$ is defined by
the ratio $(n-3)/(n-2)$, and $n$ is the power law index for the ejecta
density profile defined by $\rho \propto r^{-n}$.
$U$ is a parameter that depends on the density of the
supernova ejecta at a given ejecta speed, $\tau_{\nu,ff}$ is the
frequency dependent optical depth due to free-free absorption experienced
by the radio emission as it escapes outward from the interaction region into
the surrounding circumstellar material, and $C$ is a constant.  Initially,
the optical depth is large, depressing the emission, but as the shock moves
out, the optical depth falls with circumstellar density and the radio
emission peaks.  Note that due to the
frequency dependence of the optical depth, shorter wavelengths peak first.
According to the model, after peaking the radio emission falls
monotonically to undetectability.
The radio emission may eventually rise again either from another phase of
interaction with a denser layer of circumstellar material
(e.g., SNe~1979C, 1987A) or from the onset of the SNR phase.

How does the model apply to different SN types?
For core-collapse SNe (Type~II, Ib/c) the interaction occurs between the
outgoing shock and the pre-SN progenitor wind.  For Type~Ia SNe the
interaction can take place between the outgoing
shock front of the white dwarf explosion and the wind of a companion
red giant (\cite{bof95}, \cite{eck95}).
For detectable radio emission, there must be
enough circumstellar material (high enough $\mdot$/w) to fuel the
synchrotron emission, so for small $\mdot$/w
(of the progenitor or companion) no strong radio emission is expected.

For detected SNe, the data follow this ``standard model''
very well (e.g., \cite{wei86}, \cite{wei90}, \cite{van92}) albeit
with occasional minor modifications such as periodicities
folded into the model and interpreted (\cite{wei91})
to better fit the data.  Calculations by Chevalier (1998) indicate
that synchrotron self-absorption may be important for SNe~Ib/c and SN~1987A
as well as for SN whose progenitors have low mass-loss rates.  Ignoring
this effect may cause an overestimate of mass-loss rates.

By knowing the luminosity, the frequency and the age of an observation
we may predict
the mass-loss rate of the pre-SN progenitor provided we can determine
the other parameters.  Since most of our observations are
non-detections (i.e., upper limits to flux densities), we do not
have the data to determine all the parameters with a fitting
routine.  However, we may salvage the data by scaling the Type~II SNe data
to the parameters of the well-known SN~1979C
and scaling the Type~I SNe data to SN~1983N,
a Type Ib with similar (derived) properties to another Type Ib, SN~1984L
(This procedure is similar to one used by
\cite{wei89} to estimate (upper limits to) mass-loss rates for many other radio
SNe (see \S 3.2.1)).
Of course calculations performed in this way must be taken only as rough
estimates since every SN is not exactly like SNe~1979C or 1983N.
We are effectively forcing every SN in our data set to look like either
SN~1979C or SN~1983N, which has only limited validity, but the results
can be useful where no data were previously available.


We determined the relevant parameters for the mass-loss rate calculations
in the following manner.
A reasonable parameter for Type~II SNe is $n=20$ ($m=0.94$) (\cite{eck96}),
and the radio spectral index for SN~1979C is $\alpha=-0.72$ ($\gamma=2.4$)
from Weiler et al.\ (1986).
We may incorporate the constants $K$, $U$, and $m^{(5+\gamma)/2}$ into
one constant and determine its value by substituting all other known 
parameters and solving.  The constant $C$ can be determined
by knowing the optical depth, age, frequency and mass-loss rate
of an observation.  Weiler et al.\ (1986) have fitted data for SN~1979C
to determine their parameter $K_2$, representative of
the optical depth at $5$ GHz and $t=1$ day.  Additionally, we use the mass-loss
rate calculated by Weiler et al.\ (1986) for SN~1979C to calculate $C$.
The resulting luminosity equation is

\begin{equation}
L = 1.8 \times 10^{37} \left(\frac{\stackrel{\cdot}{M}}
{w / 10 \mbox{ km s}^{-1}}
\right)^{1.7} \left(\frac{t}{\mbox{1 day}}\right)^{-0.88}
\left(\frac{\nu}{\mbox{5 GHz}}\right)^{-0.7}
e^{-\tau_{\nu,ff}} \mbox{ erg s}^{-1}
\end{equation}
where
\begin{equation}
\tau_{\nu,ff}= 2.6 \times 10^{16}
\left(\frac{\stackrel{\cdot}{M}}{w / 10 \mbox{ km s}^{-1}} \right)^{2.18}
\left(\frac{t}{\mbox{1 day}}\right)^{-2.8}
\left(\frac{\nu}{\mbox{5 GHz}}\right)^{-2.1} .
\end{equation}

For Type~I SNe, the SN density power law index is typically $n=7$ ($m=0.8$),
and for SN~1983N, Weiler et al.\ (1986) report $\alpha = -1.03$
($\gamma = 3.1$).  Determining the other constants in the same manner for
Type II SNe results in a luminosity equation for SNe Type I

\begin{equation}
L = 9.1 \times 10^{36} \left(\frac{\stackrel{\cdot}{M}}
{w / 10 \mbox{ km s}^{-1}}
\right)^{1.4} \left(\frac{t}{\mbox{1 day}}\right)^{-1.6}
\left(\frac{\nu}{\mbox{5 GHz}}\right)^{-1.0}
e^{-\tau_{\nu,ff}} \mbox{ erg s}^{-1}
\end{equation}
where
\begin{equation}
\tau_{\nu,ff}= 3.5 \times 10^{17}
\left(\frac{\stackrel{\cdot}{M}}{w / 10 \mbox{ km s}^{-1}} \right)^{2.6}
\left(\frac{t}{\mbox{1 day}}\right)^{-2.4}
\left(\frac{\nu}{\mbox{5 GHz}}\right)^{-2.1}
\end{equation}

We take the commonly assumed value for the wind speed,
$10 \mbox{ km s}^{-1}$.
For simplicity we use this wind speed for all SN Types although we note that
for Types~Ib/c, the wind speed may be higher.
Calculated values for $\mdot$ are shown in Table~4.
In many cases the observation age has been calculated from the time of
maximum brightness or from the time of discovery since data on the
explosion date was not available, especially for the older SNe.  Thus, the ages
may be in error by about 30 days.  For any SNe (in our survey) with
observation ages older than that of
SN~1983K ($\sim$ 1500 days) this amounts to an error in the ages of less
than 2\%.  For the SNe with observation ages less than 1500 days, we were
able to obtain data on the explosion date for a better calculation
of the (upper limit to the) mass-loss rate.

For all but two of the upper limits, the observation age is large enough
that the radio emission can be assumed to be fading (for the purposes of
deriving a mass-loss rate.)  However, observations
for the Type~Ia SNe~1986G and 1989B are at 10 and 14 days past
explosion, respectively.
It is possible that the radio emission at these ages could be
rising (pre-peak) or falling (post-peak). This would result in two
different derived mass-loss rates---a larger mass-loss rate corresponding to
the optically thick case when the emission has not peaked yet and a
smaller mass-loss rate corresponding to the optically thin case when
the emission has already peaked (see \cite{eck95} for details and
previously published results for a similar case of early observations of SN~1986G).
For SNe~Ia we may be able to test whether the progenitor system is a
symbiotic (white
dwarf accreting matter from the wind of a red giant companion) using early
radio observations such as these.  By comparing the range of derived mass-loss
rates to the mass-loss rate expected from red giants in symbiotic systems
($10^{-7}$ to $10^{-5} \smpyr$ [Seaquist \& Taylor 1990,
M\"{u}rset et al. 1991]) we may test the symbiotic scenario.  Inspection
of Table~4 reveals that the derived range of mass-loss rates for SNe~1986G
and 1989B are in mild conflict with that expected for the symbiotic
red giant companion (in agreement with original calculations by Eck et al.\ [1995]
for SN~1986G).  These were probably not symbiotic systems.

\subsubsection{Model Assumptions}

We now examine some of the assumptions in the Chevalier model
and their possible effects on the derived luminosities and mass-loss rates.
One of the key assumptions in the theory of the origin of the synchrotron radio
luminosity concerns the efficiency of conversion of the thermal
energy density of electrons into the magnetic energy density
and relativistic electron energy density.
It has been assumed that this value is constant at
about $1 \%$ (\cite{che82b}), but this assumption is being questioned
based on observations of SN~1987A.  Also inherent is the assumption that
the density of the circumstellar wind material and the
outer layers of the SN ejecta are described by
time-independent power laws, $\rho_{csm} \propto r^{-2}$ and
$\rho_{SN} \propto r^{-n}$.  While the assumptions on the density profile
of the SN ejecta have been shown to be good assumptions, some cases have
shown evidence for a circumstellar power law density that looks different
from the assumed form, i.e.\ $\rho_{csm} \propto r^{-3/2}$
(SN~1993J: \cite{van94}).
This has been interpreted as a changing presupernova mass-loss rate.
Other radio SNe have shown evidence for clumpiness in the pre-SN wind or
non-spherical symmetry of the shock and a mix of internal absorbers and
emitters along the line of sight (SNe~1986J: \cite{wei90},
SN~1988Z: \cite{van93b}).  In these particular cases, the model
was modified to better fit the data.  Finally, if synchrotron
self-absorption is important, the derived upper limits to mass-loss
rates may be overestimated (\cite{che98}).  Since these are only upper limits, we do
not include these effects.

For detected SNe with enough data to form a light curve, mass-loss rates
can be derived directly.  Equation (16) from Weiler et al.\ (1986)
gives an expression
for the mass-loss rate as a function of several parameters, one of
which, $\tau_{ff}$, the free-free optical depth, can be derived by fitting the data.
Some common assumptions on the other parameters are $v_{shock} =
10^4 \mbox{ km s}^{-1}, v_{wind} = 10 \mbox{ km s}^{-1},
\mbox{ and } T_{e} = 10^4$ K in the wind.  How sensitive are the
mass-loss rates to these values?  While the functional form for the
dependence of these parameters on the mass-loss rate is
obvious from the expression from Weiler et al.\ (1986), there are
other considerations.  For example, Lundqvist \& Fransson (1988)
report that the assumption of a fully ionized wind may
underestimate the mass-loss rate by a factor of 2 or more.
To review this and some other considerations affecting the radio emission,
we refer the reader to Lundqvist \& Fransson (1988).

In the calculation of the (upper limits to) mass-loss rates for Type~I SNe,
we have assumed that the Type~Ib/c and Type~Ia SNe share model
parameters.  Despite the lack of detections of Type Ia SNe, we have followed
Weiler et al.\ (1989) and performed calculations using this assumption in an effort
to salvage our data on Type~Ia SNe.

\subsection{Comparisons}

\subsubsection{Other Radio SN Searches}

Montes et al.\ (1997) reported a detection of SN~1986E at an
age of about 8 months.  By scaling some of the parameters to those
of SNe~1979C \& 1980K and using the detection of SN~1986E as well as
some tight upper limits, they derive a
light curve and a subsequent mass-loss rate of $> 4.7 \times 10^{-5}
\smpyr$.  While our upper limits are consistent with their radio
light curve (i.e., their derived radio light curve predicts the radio emission
to be less than our upper limits at the epochs of our observation), our
derived mass-loss rate appears to be in conflict, at $< 4.9 \times 10^{-6} \smpyr$,
by an order of magnitude.  However, there are several items to be noted in 
comparing the results of Montes et al.\ and the results found here.  
First, Montes et al.\ apparently derive the mass-loss rate using only the 
free-free absorption term (equation 2, here) while a somewhat different 
form has been used in our calculation, i.e., both equations 1 and 2.  
It is also noted that for most of the epochs of our observations, the 
mass-loss rate derives mainly from equation 1 since it is probable that 
optical depth is very small.  
Second, by using the Montes et al.\ detection and our 
parameters for Type II SNe, we find two mass-loss rates: 
one corresponding to the
optically thick case ($6.5 \times 10^{-5}  \smpyr$) and one 
corresponding to the 
optically thin case ($4.1 \times 10^{-6}  \smpyr$).  
It is unclear whether SN~1986E
is pre-peak (optically thick) or post-peak (optically thin) at the epoch of the 
Montes et al.\ observation and therefore unclear whether there is a 
conflict between 
our results and Montes et al.  A calculation using the Montes et al. upper 
limits 
in the same manner restricts the mass-loss rates to be 
$2.1 \times 10^{-6} \smpyr < \  \mdot \ < 1.2 \times 10^{-5} \smpyr$ at the earliest epochs 
($< 200$ days) and 
$\mdot < 4.4 \times 10^{-6} \smpyr$ at the later epochs 
($> \sim 200$ days).  
Third, the epochs of Montes et al.\ observations and ours are 
separated by $\sim 6 - 8$
years.  (It is unlikely that SN~1986E is still in an optically thick 
phase at the 
epochs of our observation since unreasonably high mass loss rates ($> 10^{-3} 
\smpyr$) 
are required using our model parameters.  
Only the optically thin result has been quoted 
here).  

Our upper limit to the pre-SN mass-loss rate 
for SN~1986E is consistent with the Montes et al.\ observation 
if the SN was in an
optically thin phase at both epochs.  If the SN was in an optically thick phase
at the Montes et al.\ epoch, there 
appears to be a conflict between our model 
results and theirs, however there is some 
evidence for a changing pre-SN mass-loss rate in other SNe 
(e.g., SN 1993J \cite{van94}).  

To analyze the sensitivity of the calculation to the model 
parameters, we calculate 
the mass-loss rate for all of the upper limits and observations of SN~1986E by 
scaling to the Montes parameters and by scaling to SN~1979C.  For the same observation,
the (upper limits to) 
mass-loss rates are lower by a factor of $5 - 10$ by scaling to 
SN~1979C versus scaling
to SN~1986E.  This indicates that the assumptions on parameters 
are important and that
our results may only be accurate to within an order of 
magnitude, but again our results
can be useful where no other data is available.


Brown \& Marscher (1978) report radio upper limits to 46 SNe at two wavelengths
and at ages varying from a few months to more than 70 years.
Their sample contains many of the SNe in the present sample (SNe~1895B,
1909A, 1921B, 1921C, 1937F, 1951H, 1954J, 1959D, 1961V, 1969L,
1970G, 1972E and 1973R),
but with relatively high upper limits (by today's standards)
in the range $3-30$ mJy; they reported no detections.
By applying our models to their data on upper limits,
we may derive upper limits to mass-loss rates similarly to what we have done
with our own data.   Despite some early observations
from Brown \& Marscher at less than a year after explosion, no upper limits
to mass-loss rates were obtained that are lower than ours.

Weiler et al.\ (1989) first performed a procedure very similar to what we have
done in calculating upper limits to mass-loss rates for 24 SNe observed
by them at 6 cm, including SNe~I and II.  They calculate mass-loss rates
by scaling all Type~II to SNe~1979C \& 1980K and all Type~I to SN~1983N, but
they do not include absorption effects (except for the fluxes from
SNe~1979C \& 1980K).  At late times absorption is
negligible, but it can be important at earlier epochs.  By not including
an absorption term, this could cause an underestimate in mass-loss rates.
We calculate mass-loss rates for the 24 SNe in Weiler et al.\ (1989)
and derive upper limits
that are $10\% - 60\%$ greater than theirs for their Type~II SNe.
For the Type~I SNe in Weiler et al.\ (1989), including absorption effects
depresses the emission enough that any physically reasonable mass-loss rate
would have produced emission below the flux density upper limits for many of
their older, more distant SNe.  For the closer Type~I SNe observed early enough
to restrict mass-loss rates, our upper limits to the mass-loss rates are
greater than theirs.  The $10\% - 60\%$ differences in the results from
applying our model to Weiler et al.\ upper limits are likely mainly due to free-free
absorption effects, but could also be due to parameter differences.

\subsubsection{Radio and optical emission}

Can radio emission be predicted?  Since radio emission is the result of
shock-circumstellar matter interaction (radio SNe) or shock-interstellar
matter (radio SNR)
interaction, any evidence for circumstellar/interstellar interaction
(in other wavebands) at SNe sites might be used as a predictor for radio emission.

Since most SNe have faded in the optical by an age of about 2 years,
indications of optical emission beyond 2 years may be the result
of some type of enhanced circumstellar interaction.
There are 12 SNe that have been detected in the optical at an age of
greater than 2 years.  Of these, SN~1885A is the only Type~Ia, with optical
detection at 103 yrs (Fesen, Hamilton, \& Saken 1989,
\cite{fes97}).  However, since SN~1885A has been detected in the optical via
absorption, there is no evidence for circumstellar interaction, so we remove it
from our group of 12.   All other SNe with evidence for optical emission at ages
greater than 2 years are core-collapse SNe: SN~1957D at 30-32 yrs
(Long, Blair, \& Krzeminski 1989),
SN~1961V at 22-24 yrs (\cite{goo89}),
SN~1970G at about 22 yrs (\cite{fes93a}),
SN~1978K at 12-20 years (\cite{ryd93}, Chugai, Danziger \& Della
Valle 1995; Chu et al. 1999, \cite{sch99}),
SN~1979C at 10 yrs (\cite{fes93}),
SN~1980K at 7-9 yrs (\cite{fes90}),
SN~1985L at 12  yrs (\cite{fes98}),
SN~1986E at 7-8 yrs (Cappellaro, Danziger, \& Turatto 1995),
SN~1986J at 4-7 yrs (\cite{lei91}), 
SN~1987A at about 10 years (\cite{son98}), 
and SN~1988Z up to 8.5 years (\cite{are99}).  Some of these SNe with optical
recoveries have been observed to be radio emitters.
SN~1987A has been detected in the radio at an age of about 10 yrs but at
a flux level that would be undetectable at much greater distances.
SNe~1957D and 1961V were recovered in the optical only after detection
in the radio. (It is not clear whether SN~1961V was an SN or an LBV
outburst [\cite{goo89}, \cite{fil95}], although recent radio observations
support a supernova interpretation Stockdale et al. 2001b).
SN~1970G was optically recovered shortly after the last radio detection
(\cite{cow91}) which indicated a steep decline rate in the radio when combined
with initial radio observations (\cite{got72}, \cite{all76}).
Schlegel et al. (1999) report fading radio emission from SN~1978K since 1992 and report observations that indicate the optical light curve approaching a constant 
apparent magnitude starting at about 10 years past explosion. 
Observations of SN~1979C indicated fading radio emission
(albeit with approximate sinusoidal variations [\cite{wei92b}])
at the time of
optical recovery, but more recent observations indicate that it is
now brightening in the radio (\cite{mon00}).
SN~1980K was optically recovered when the radio emission was fading, and
recent radio results now indicate a sharper drop than
previously reported (\cite{mon98}) despite continued optical detections
(Fesen, Hurford, \& Matonick 1995).
There is no available radio data SN~1985L at an age of greater than 3 yrs, but
6 cm data indicated fading emission near that time (\cite{van98}).
Although still fading from its initial rise,
SN~1986J was emitting strongly in the radio at an age of 6 yrs.
SNe~1885A (Crane, Dickel, \& Cowan 1992) and 1986E (\cite{eck96})
do not show detectable radio emission near the
time of optical recovery although SN~1986E did show one radio detection
at an age of 8 months (\cite{mon97}).  Observations of SN~1988Z show strong radio emission at an age of about 5 years (\cite{van93b}).  Model fits to the data at 5 years imply post-peak radio emission at 2, 3.6, and 6 cm but not at 
20 cm (\cite{van93b}).  The optical light curve for SN~1988Z shows 
strong but generally fading optical emission (\cite{are99}) in this same epoch.

To summarize: of all 11 SNe (without SN~1885A) with optical recoveries, 9 show
radio emission.
Of these 9, only two (SNe~1979C and 1987A) show a rise
in radio emission
within a few years after optical recovery.  It is unlikely that SNe~1979C
and 1987A are entering a radio SNR phase at an age of only 10-20 yrs.
Thus, optical recovery precedes an increase in radio
emission while still in a RSN phase only for SNe~1979C and 1987A.
Is optical recovery a good indicator (in general) of radio emission as the
SN shock encounters circumstellar material?
For a majority of the cases (9 out of 11), it would seem to be
true.
Support for a general correlation between late-time optical and radio
emission is found in the behavior of SN 1957D, which showed a dramatic
drop in the optical (Long, Winkler \& Blair 1992)  and contemporaneously
in the radio (Cowan et al. 1994). A similar correlation 
was seen for SNe~1979C and 1980K between 
a leveling (SN~1979C) or drop (SN~1980K) 
in radio (Montes et al. 2000, 1998) and the behavior of the 
optical emission (Fesen et al. 1999). 
Although both radio and late-time optical emission (e.g., see 
\cite{fes99} for discussion) may be indicative of
enhanced circumstellar interaction, we caution that the emission mechanisms are
different, as may be the regions where the emission originates.
Clearly, further combined radio and optical studies will be required to
understand better the long-term  radio and optical correlations.

\subsection{Luminosity-Age Plots}

Figures 1 -- 5 are plots of the radio luminosities for ($3\sigma$)
upper limits versus
age of observation for all of the SNe in our sample.  The plots are
separated by SN Type and wavelength.  Each plot shows upper limits
(as upside-down triangles), detections (larger unfilled polygons),
and at least one SN of the same type with its model light curve
(solid and dashed curves) and observational data (smaller unfilled
circles).  Because of its unknown SN type, SN~1945B (\cite{lil90}) was
included on both Type~I and II SN plots.
Although SNe~1950B and 1957D are both of unknown type, they are likely to
be Type~II SNe (\cite{cow94}, \cite{eck98}) and were
included only in the Type~II plots.

\subsubsection{Type II SNe}

Inspection of the Type~II plots at both 20 and 6 cm reveal that
upper limits for SNe~1980D, 1984E \& 1986E
are below the data for SN~1979C but above the data for
SN~1980K.  Given that
the radio emission from SN~1980K has fallen sharply,
it is possible that most of the other upper limits (for the older SNe)
have fluxes above that of SN~1980K.  Recent data from Montes et al.\
(1998) indicate that radio emission from SN~1979C is now rising again 
instead of continuing to fade.  The inconsistency of our upper limits and 
detections with the SN~1979C data and possible similarities with the 
SN~1980K data indicate that the Type~II SNe in our survey
have properties more like SN~1980K than SN~1979C.  
It also suggests that SN~1980K may be a more typical Type~II radio 
emitter than SN~1979C.

\subsubsection{Type I SNe}

The observation age of SN~1895B makes it interesting to consider.  We
may compare it with model curves (dashed curves) for the radio turn-on
of a SNR as
calculated by Cowsik \& Sarkar (1984).  The dashed curves correspond to
their models c) (constant density piston) and d) (isothermal piston) and
assume an explosion energy of $10^{51}$ ergs, $0.5$ solar masses of
ejected material, an interstellar density of 1 baryon cm$^{-1}$ and
a spectral index, $\alpha \simeq -1$.  It is clear that SN~1895B
{\em could} be in the SNR phase, but the high upper limit
does not exclude the possibility that it may still be fading as an RSN.
At 6 cm, our data is very limited for Type~I SNe.
Since no Type~Ia SNe have been detected in the
radio, we cannot conclude much about the comparison of the upper limits
and the SN~1983N light curve, except that our upper limits are below the
light curve.
For the 3.6 cm plot, no data was available for SN~1983N so we plot data
and curves from the only other Type~I SN with 3.6 cm data, the Type~Ic SN~1990B.
Without other radio light curves to compare our upper limits to,
it is difficult to say anything with confidence about the upper limits
(for the Type Ia SNe~1885A and 1989B) except that they are
in mild conflict with the SN~1990B data.

\section{CONCLUSIONS}
We summarize the results of the analysis of the radio observations in our
survey.  Many of the conclusions found from analysis of our observations
agree with and confirm
conclusions found in other radio SNe papers (e.\ g.\ , Weiler et al.\ 1989).
These observations increase the size of the data pool on radio observations of
SNe and provide more data with which to test theories of radio emission
from SNe.

\begin{enumerate}
\item Most SNe are not detectable radio emitters with current radio
telescopes due to either large distances or intrinsically faint
radio supernovae (or early enough observations).

\item No Type~Ia SNe have been detected despite some early and/or deep observations
(SNe~1885A [\cite{cra92}], 1895B, 1937C, 1986G, 1989B).

\item While most of the Type~II SNe observed in our survey were
{\em not detected} in the radio, we find that most of the {\em detected}
ones (of all the SN types) are Type~II SNe.

\item Assuming the Chevalier model, our flux upper limits imply
a lack of circumstellar material (or fuel) for detectable radio emission.  Radio
emission is thus dependent on pre-SN mass-loss and on mass-loss history.

\item Using the Chevalier model for Type~Ia SNe, one scenario for which
we could get radio emission is the symbiotic-star progenitor
scenario (\cite{bof95}).
Our non-detections of SNe~1986G and 1989B indicate that the derived
upper limits to mass-loss rates are in mild conflict with the mass-loss
rates inferred from red giants in symbiotic systems.  These SNe are probably
not symbiotics.  The other Type~Ia SNe (SNe~1885A [Crane et al.\ 1992],
1895B, 1937C) have higher upper
limits to the mass-loss rates so that we cannot discount the symbiotic-star
scenario as a progenitor scenario.

\item By applying the Chevalier model to our Type~II SNe observations, we see that
no Type~II SNe have been detected with a derived value for the mass-loss
rate of $\mdot \simlt 10^{-6} \smpyr$, implying a lack of
circumstellar material.  This value is very dependent on the
scaling of the parameters of our SNe to those of the well-observed
SN~1979C, and the uncertainty in our derived mass-loss rates may be as high
as a factor of 10 due to dependencies on parameters (see SN~1986E).

\item The derived mass-loss rates can also have a dependence on the
assumptions inherent in the Chevalier model.  The effects of these assumptions
on the radio luminosity are not trivial to calculate and warrant further
study (e.g., see Lundqvist \& Fransson [1988] for an investigation of some of these
effects).

\item There are 11 SNe with optical recoveries (indicating
circumstellar interaction) at an age of
greater than 4 years.  Of the SNe with optical recoveries, 9 have been
detected as radio emitters (and for 2 cases of these 9, the optical emission
preceded a second rise of radio emission).  This implies that
radio emission may be associated with optical recovery.

\item Since the upper limits for Type~II SNe in our survey are more consistent
with SN~1980K than with SN~1979C, these Type~II SNe in our survey may have radio
properties more similar to SN~1980K than SN~1979C.

Finally, we note that despite more than two decades of theoretical and
observational work in radio supernovae, radio emission from supernovae is
not completely understood.  As the intermediate-age radio
supernovae are followed, we hope to explore further the
transition of radio supernovae into supernova remnants.
More sensitive observations of young and
intermediate-age SNe, as well as more theoretical work, are needed to help
us fully understand the origin and evolution of RSNe.
\end{enumerate}

\acknowledgments We would like to thank the anonymous referees for
helpful comments and suggestions, Schuyler Van Dyk for sharing
his optical H$\alpha$ maps with us, Paul Hodge for providing positions
of H~II regions, Bill Romanishin for observing and reducing optical data on
NGC 3627 for us, Kurt Weiler for helpful discussions,
W. Miller Goss for providing the raw data on SN 1951H, Pat Crane for the
upper limit on SN 1885A and Doug Roberts for help
on the data analysis using AIPS.  This work has been supported
in part by NSF grants AST-9986974  and AST-9618332
at the University of Oklahoma.

This research has made use of the NASA/IPAC Extragalactic Database (NED)
which is operated by the Jet Propulsion Laboratory, California Institute
of Technology, under contract with the National Aeronautics and Space
Administration.


\section*{Appendix A.  Non-Supernova Radio Sources}

To ensure completeness, but not take away from the focus of the paper,
we have included this appendix as a description of the analysis of the non-SN
radio sources in our maps.  The non-supernova radio sources detected in our
observations are listed in Table~5.  Selected maps with more than
3 sources can be found in Figures~6 to 9.  All peak fluxes and
positions were
determined by using AIPS (Astronomical Image Processing System provided by the
NRAO) to fit the source to a two-dimensional Gaussian on top
of a linearly sloping background.  The fit to sources $\epsilon$
and $\pi$ in NGC~3627 failed to report an uncertainty in flux,
so we report the rms noise as representative of the flux uncertainty.

Many sources have been identified as H~II regions via superposition
onto H$\alpha$ maps or via comparison with published positions of H~II regions,
but some caution must be exercised.  Without a spectral index to verify the
thermal nature of the radiation from these sources, we can only tentatively
identify the radio emission as originating in the H~II regions.
(Consider the case of SN~1957D whose emission was initially
non-thermal.  Later measurements indicated that the emission is thermal and
is believed to have faded below that of an associated H~II region
[\cite{cow94}].)

\paragraph{NGC 1058}
Radio detection of SN~1961V has been reported by Cowan et al. (1988),
Cowan \& Branch (1985), and Cowan et al.\ (1991).  An H$\alpha$ map
from Cowan et al.\ (1988) was available for comparison with our sources, but
$\alpha$ had a position outside ($> 4\amin$ from galaxy center)
the H$\alpha$ field of view.  On the other hand, $\beta$ was within
the field of view for the H$\alpha$ map but did not correspond to any
H$\alpha$ emission region.

\paragraph{NGC 2403}
A contour map of the radio sources in NGC~2403 can be seen in Figure~6.
We detected 15 sources with a signal-to-noise of $5\sigma$ or greater.
Turner \& Ho (1994) report 6 cm radio observations of NGC~2403,
including 7 sources whose positions lie within $1\asec$ to $5\asec$
of a corresponding source found in our 20 cm map.  However, spectral
index calculations were not possible since none of the mutually observed
sources had the same beamsize and none were unresolved.  Examination
of integrated fluxes reveals that source $\beta$ is consistent with
thermal emission.

A map of H~II regions from Sivan et al.\ (1990) superimposed onto
our radio map reveals that $\beta$ lies directly over a group
of H~II regions (including No.\ 293 in \cite{siv90}).  Source
$\beta$ is likely a group of H~II regions.  The same positional coincidence
with a radio source and an H~II region(s) is found for sources $\gamma$,
$\delta$, $\epsilon$, $\rho$, $\theta$, $\omega$, $\eta$, $\chi$ and
$\xi$.  They are likely H~II regions as well.   We find that the
luminosities of these sources is comparable to H~II regions identified
in Cowan et al.\ (1994).

Comparison with reported SNRs in NGC~2403 by Matonick \& Fesen (1997) reveals
that source $\mu$ has a peak position very close to that of SNRs No.\ 6 \& 7.
At a distance of 3.2 Mpc to NGC 2403 (see Table 2), $\mu$ is about 50 pc
from both SNRs No. 6 \& 7.  Since SNR No. 7 has a reported diameter of 60 pc,
it is very possible that $\mu$ is identified with this SNR.  More radio
observations are recommended to confirm this identification.

\paragraph{NGC 3169}
The first source in the list has a position that is coincident with the
central regions of NGC~3169.  The emission is probably associated with the
center of the galaxy.

\paragraph{NGC 3627}
Prior observations of NGC~3627 were found in Hummel et al. (1987) (1.49 GHz),
Crane (1977) (2.7 \& 8.1 GHz), Zhang, Wright, \& Alexander (1993)
(CO and H I) and Hodge (1974) (optical H~II).
Source $\alpha$ has a non-thermal spectral index and a position within
the uncertainty ($10\asec$) for the galactic center.
It lies directly over the CO peak
in the center of the galaxy from Zhang et al. (1993)
and is probably associated with the galactic center.
Hummel et al. observe a source $< 0\fasec 5$ from $\alpha$ with an integrated
flux over the inner $2\asec$ of $15 \pm 2$ mJy.  We measure the integrated
flux over approximately the same region to be $12.5 \pm 0.1$ mJy.

H$\alpha$ observations taken on the 18-inch telescope at University
of Oklahoma by W. Romanishin (1997) reveal an emission region lying over
$\beta$.

Hodge (1974) reports observations of H~II regions expressed as offsets
and plots them relative to the galactic center.  If we superpose
$\alpha$ with the position for the galactic center on the plot,
we find that $\beta$ lies directly over two H~II regions (No.\ 12 \&
13 in Hodge) indicating that $\beta$ may be an H~II region.
Source $\gamma$ has no counterpart in Hodge (1974).

Source $\epsilon$ has a position that is coincident with the position of
an H~II region (No.\ 47) reported by Hodge (1974).  This coincidence
assumes that the center of the galaxy is at $\alpha$.  Examination of
the fluxes at two wavelengths indicates that the emission from $\epsilon$
is non-thermal.  This probably indicates that there is an SNR nearby
or possibly associated with the H~II region.  If we again assume
that $\alpha$ is the center of the galaxy, we find that $\pi$ lies
$4\asec$ from the position of an H~II region (No.\ 50), so it is not clear
whether $\pi$ is associated with an H~II region or not.

\paragraph{NGC 4214}
A contour map of most of the sources in this galaxy can be seen in Figure~7.
There were at least 8 radio sources in this irregular galaxy with one ($\eta$)
apparently composed of a number of smaller sources.  Using an H$\alpha$
map of the galaxy provided by Van Dyk (1997) and superposing our radio
map with it, we can attempt to identify probable H~II regions.  Sources
$\alpha$, $\beta$, $\gamma$, $\epsilon$ and $\eta$ have excellent positional
coincidences with H$\alpha$ emitting regions and are all probably H~II regions.
Hartmann, Geller, \& Huchra (1986) list positions of H~II regions,
and their source H59 lies $3\asec$ from the peak position of source $\rho$.
This source is likely to be an H~II region.  In addition, the Hartmann et al.
(1986) source H48 lies $3\asec$ from the strongest source in the $\eta$
complex.  H54 \& H55 lie close
($5\fasec 5$ and $2\asec$, respectively) to the peak positions of $\beta$ and
$\alpha$ to make their identification probable.

\paragraph{NGC 4302}
Analysis of the sources in this galaxy were originally reported in
Eck et al. (1996).

\paragraph{NGC 4688}
Superposition of the Digital Sky Survey (DSS) R-band image onto our radio map
reveals no source coincident with the only prominent radio feature in our map.

\paragraph{NGC 5128}
We report measurements of the radio bright center of NGC~5128 (Cen A) at 6 and
2 cm.  The integrated fluxes are $6.1$ Jy and $0.38$ Jy at 6 and 2 cm,
respectively and the resulting spectral index is $\alpha^6_2 = -2.5$.

\paragraph{NGC 5457 (M 101), S}
A map of the (southern) region surrounding NGC~5455 and containing
many of the sources can be seen in Figure 8.  SN~1970G is visible as
a prominence pointing to the northwest from $\alpha$.
Source $\alpha$ has a flat spectral index and an excellent positional 
coincidence with NGC~5455, a known H~II region.  Sources $\delta$ and
$\omega$ have spectral indices between $0.0$ and $0.1$ indicating they are
likely H~II regions as well.  If we compare the positions of our sources
with those of H~II regions found by Hodge et al. (1990), we find positional
coincidences (within the uncertainties) for $\psi$, $\rho$, $\theta$,
$\epsilon$ and $\delta$.  These are also probable H~II regions.
Comparison of the positions of SNRs
in M~101 reported by Matonick \& Fesen (1997) with our sources reveals no
positional coincidences.

\paragraph{NGC 5457 (M 101), E}
A map of the eastern region of M~101 (containing the site of SN~1951H)
can be seen in Figure~9. Prominent
features include emission regions from the H~II regions NGC~5461 \& 5462.
Superposition of the radio map with H$\alpha$ maps found in Israel, Goss,
\& Allen (1975) reveals that sources $\kappa$, $\eta$, $\phi$, and $\chi$
in NGC~5461
lie over H$\alpha$ emission regions.  The same positional coincidence is found
for $\gamma$, $\epsilon$, $\delta$, $\beta$ and $\alpha$ in NGC~5462.  All
of these are probable H~II regions.  The positions of our sources can be
compared to H~II regions observed by Hodge et al. (1990).  We find that
$\pi$, $\eta$, $\phi$, $\gamma$, $\epsilon$ and $\alpha$ have peak
positions that are within the uncertainties for the H~II regions.
In addition, $\lambda$, $\kappa$, $\chi$, $\delta$, $\beta$ and $\theta$
have peak positions within twice the uncertainties of a particular H~II
region.  The Hodge et al. (1990) data indicate that these sources {\em may}
be H~II regions.

Both comparisons indicate that $\kappa$, $\eta$, $\phi$, $\chi$, $\pi$,
$\gamma$, $\epsilon$, $\delta$, $\beta$ and $\alpha$ are all probable
H~II regions.

Comparison with SNRs in M~101 from Matonick \& Fesen (1997) shows no
SNRs coincident with positions of any of our sources in the eastern region
of M~101.

\paragraph{IC 4182}
There are two prominent sources ($\alpha$ and $\beta$) in the west side
of the radio map.  Comparison with the VLA survey Faint Images of the Radio
Sky and Twenty-centimeters (FIRST) [Becker, White, \& Helfand 1995]
reveals two sources in that survey with peak positions
within $2\fasec 5$ and $1\fasec 0$ of $\alpha$ and $\beta$, respectively.
These are very likely to be the same sources.  Willis, Oosterbaan, \&
de Ruiter (1976) conducted a Westerbork Synthesis Radio Telescope (WSRT)
survey including IC~4182 and observe two sources whose positions are
within $3\asec$ and $11\asec$ of $\alpha$ and $\beta$, respectively, although
the peak fluxes are somewhat larger.

\clearpage
\pagestyle{empty}

\newpage
\begin{figure}
\epsscale{0.9}
\vskip -.7truein
\hskip -.5truein \plotone{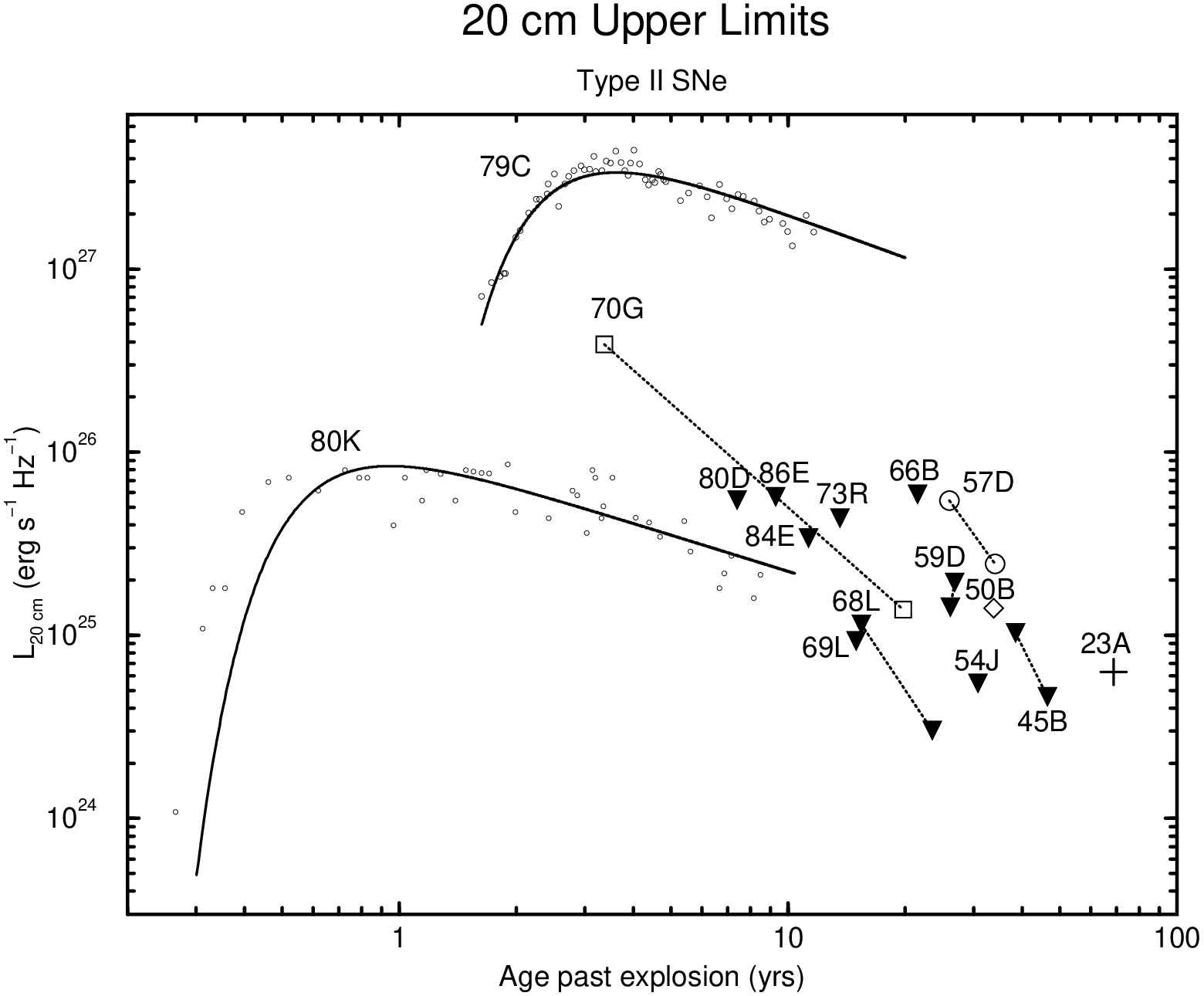}
\vskip -.5truein
\caption{
A plot of luminosity versus age for the SNe~II in our survey
at 20~cm.
Our upper limits are designated as upside-down triangles, the unfilled
shapes (squares for SN~1970G, larger circles for SN~1957D,
diamond for SN~1950B, plus sign for SN~1923A)
are data for the radio detections in our survey.
Included are data (small unfilled circles) and model fits (solid curves)
to the data for two well-observed SNe~II, SNe~1979C
(Weiler et al.\ [1986,1991]) and 1980K (Weiler et al.\ [1986,1992a]).
\label{fig1} }
\end{figure}

\newpage
\begin{figure}
\epsscale{0.7}
\hskip -.7truein \plotone{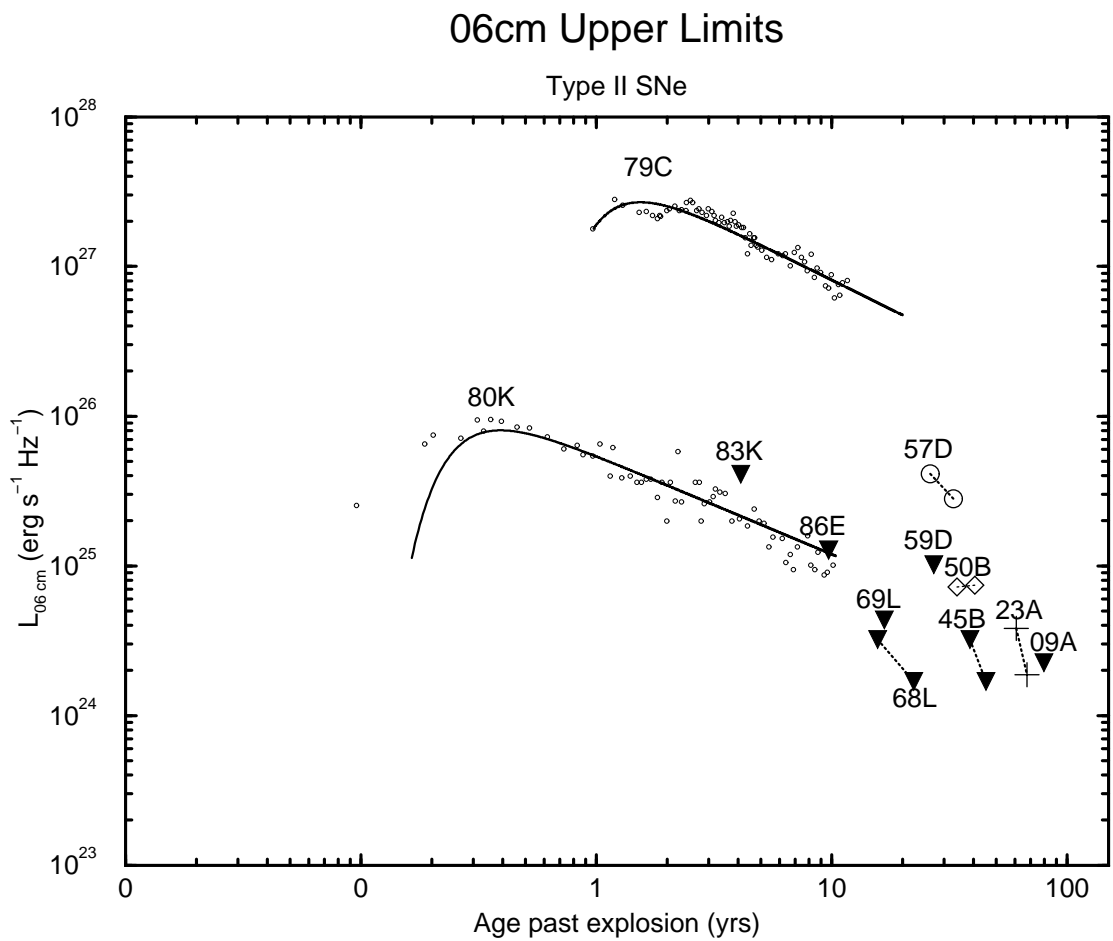}
\caption{
A plot of luminosity versus age for the SNe~II in our survey
at 6~cm. The data and curves are labeled as in Figure~1.  The data and
light curves for SNe~1979C and 1980K are from Weiler et al.\ (1986, 1991)
and Weiler et al. (1986,1992), respectively.
\label{fig2} }
\end{figure}

\newpage
\begin{figure}
\epsscale{0.7}
\hskip -.7truein \plotone{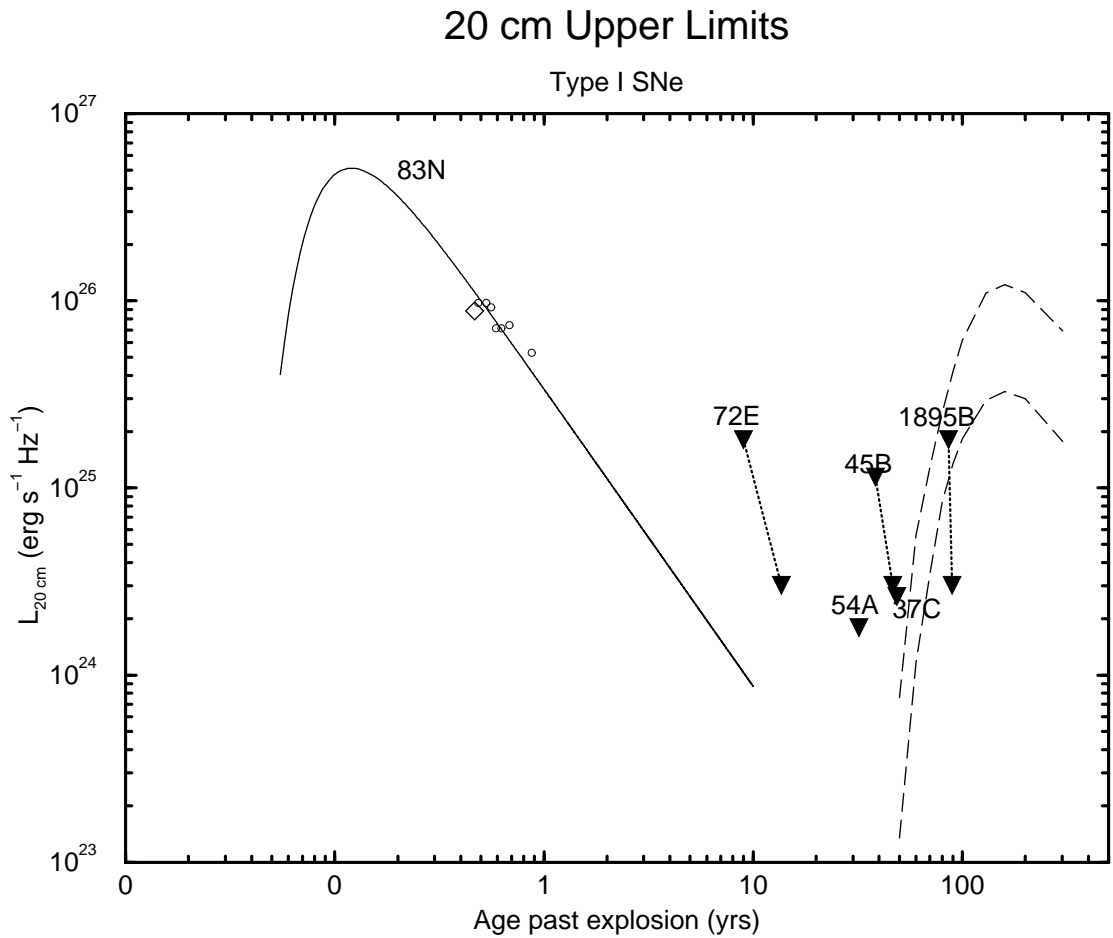}
\caption{
A plot of luminosity versus age for the Type~I SNe in our survey
at 20~cm.  Our upper limits are designated as upside-down triangles and
the diamond represents our datum for SN~1983N.  The smaller unfilled
circles and the solid curve are data and a model fit to the data
from Weiler et al. (1986) for SN~1983N.  The dashed curves are
light curves for the onset of the SNR phase from calculations by
Cowsik \& Sarkar (1984) based on the Gull (1973) model.  The two
curves are for different model assumptions (see \S 3.3.2).
\label{fig3} }
\end{figure}

\newpage
\begin{figure}
\epsscale{0.7}
\hskip -.7truein \plotone{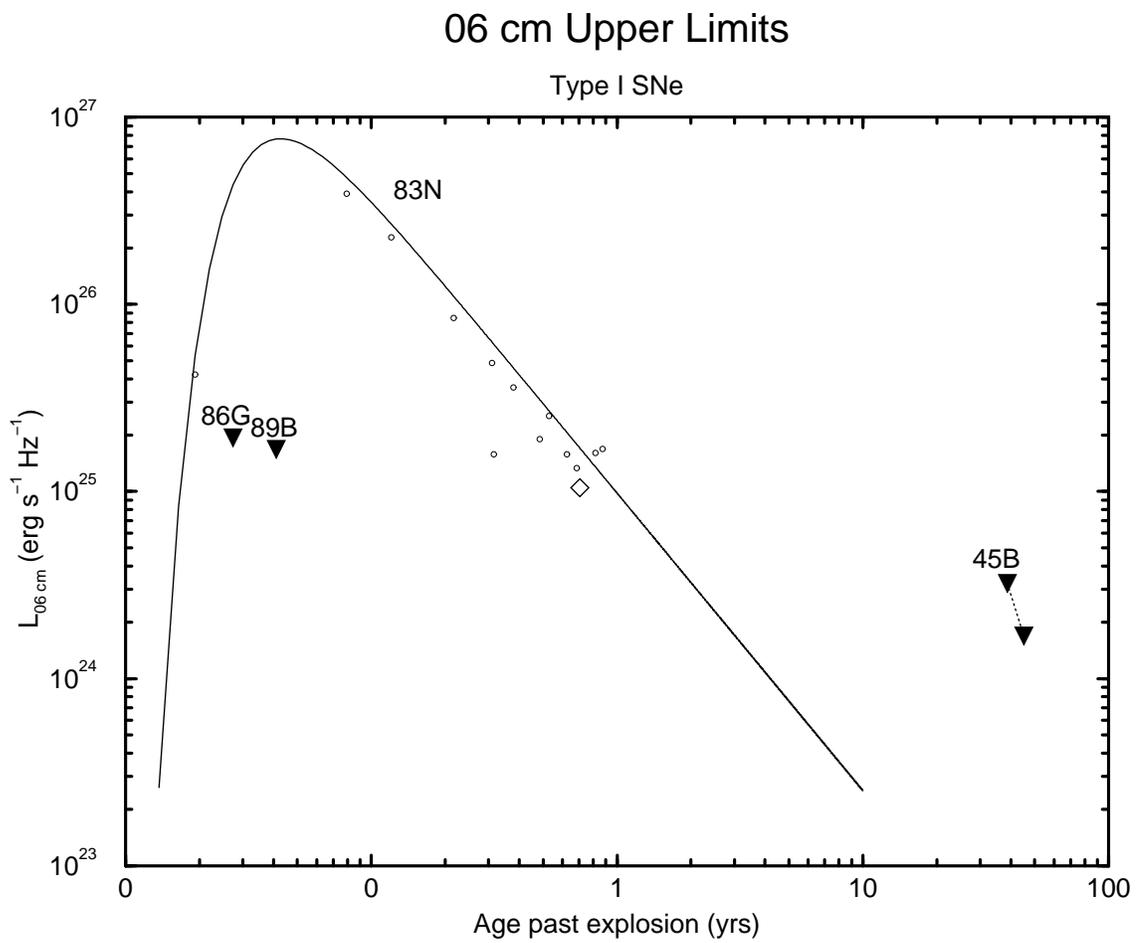}
\caption{
A plot of luminosity versus age for the Type~I SNe in our survey
at 6~cm.  Designations are as in Figure~3.
\label{fig4} }
\end{figure}

\newpage
\begin{figure}
\epsscale{0.7}
\hskip -.7truein \plotone{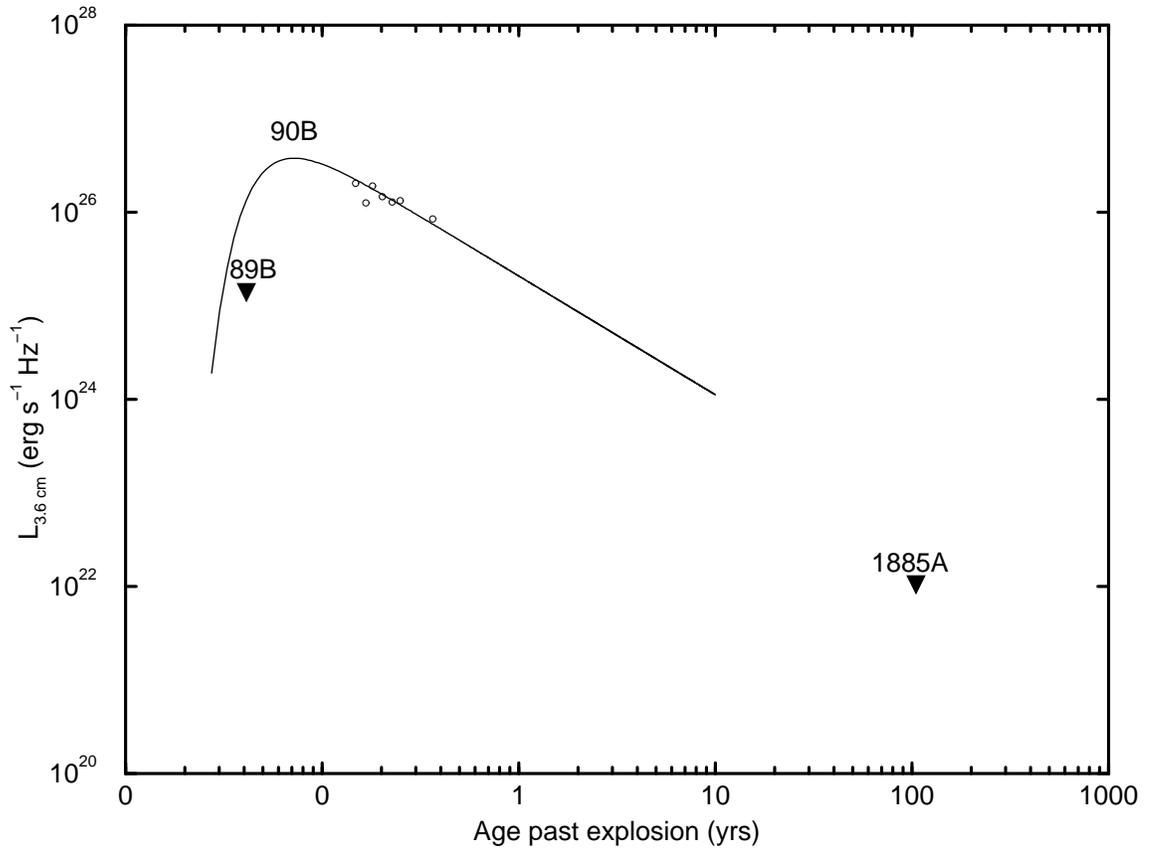}
\caption{
A plot of luminosity versus age for the Type~I SNe in our survey
at 3.6~cm. Upper limits are designated as upside-down triangles.
The upper limit for SN~1885A is from Crane et al. (1992).  Data (small
circles) and light curve (solid curve) are from the Type~Ic SN~1990B
(Van Dyk et al.\ 1993a) since data at 3.6 cm was unavailable for SN~1983N.
\label{fig5} }
\end{figure}

\newpage
\begin{figure}
\plotone{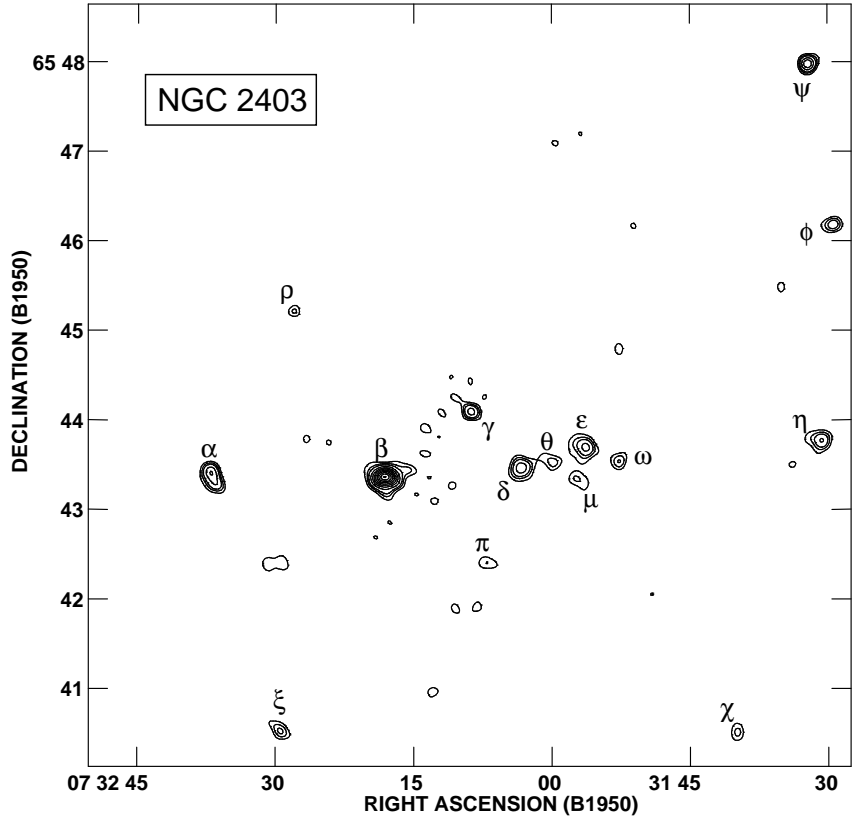}
\caption{
A contour map of the field of view for the 15 sources in NGC~2403
at 20~cm, labeled as in Table~5.  Sources $\beta$, $\gamma$, $\delta$,
$\epsilon$, $\rho$, $\theta$, $\omega$, $\eta$, $\chi$ and $\xi$ are
probably H~II regions (see text).  Contour levels are at $0.47$, $0.70$,
$0.94$, $1.4$, $1.9$, $2.3$, $2.8$, $3.3$, $3.7$, $4.2$, and $4.6$
mJy~beam$^{-1}$ and the rms noise is $0.12$ mJy~beam$^{-1}$.
\label{fig6} }
\end{figure}

\newpage
\begin{figure}
\plotone{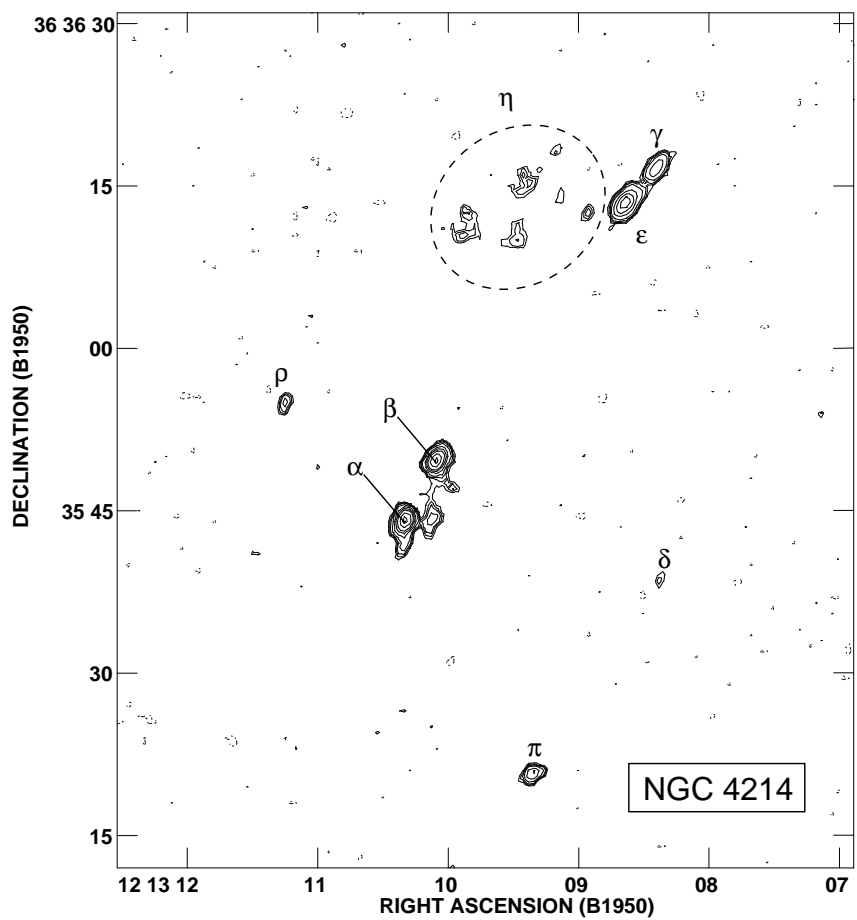}
\caption{
A contour map of part of the field of view for NGC~4214 at
20~cm, labeled as in Table~5.  The sources that appear
to be a part of a larger complex are circled and labeled $\eta$.
Sources $\alpha$, $\beta$, $\gamma$ and $\eta$ are all likely to be
H~II regions (see text).  Contour levels are at
$-0.053$, $0.066$, $0.083$, $0.099$, $0.13$, $0.20$, $0.27$, $0.33$,
$0.47$, $0.60$ and $0.66$ mJy~beam$^{-1}$ and the rms noise is $0.022$
mJy~beam$^{-1}$.
\label{fig7} }
\end{figure}

\newpage
\begin{figure}
\plotone{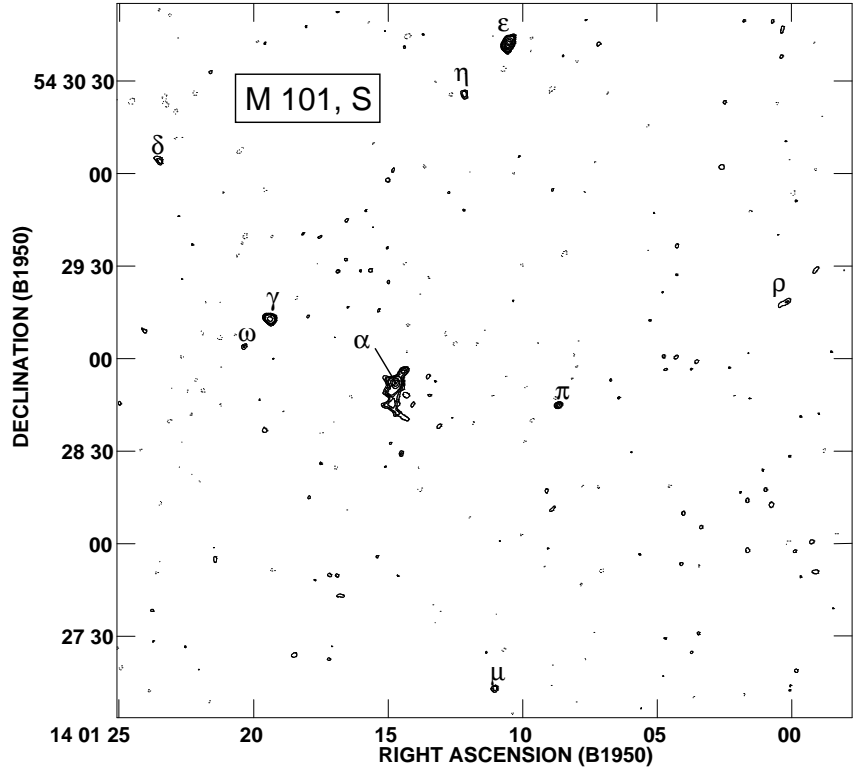}
\caption{
A contour map of part of the field of view for the southern
region of NGC~5457 (M~101) at 20~cm, labeled as in Table~5.  Emission from
SN~1970G (source $\beta$) is visible just to the northwest of $\alpha$, but
not labeled. Sources $\alpha$, $\delta$ and $\omega$ are probably H~II regions
from their flat spectral indices (see Table 5).  Contour levels are at
$-0.065$, $0.065$, $0.093$, $0.14$, $0.19$, $0.37$, $0.56$, $0.74$ and $0.92$
mJy~beam$^{-1}$ and the rms noise is $0.021$ mJy~beam$^{-1}$.
\label{fig8} }
\end{figure}

\newpage
\begin{figure}
\plotone{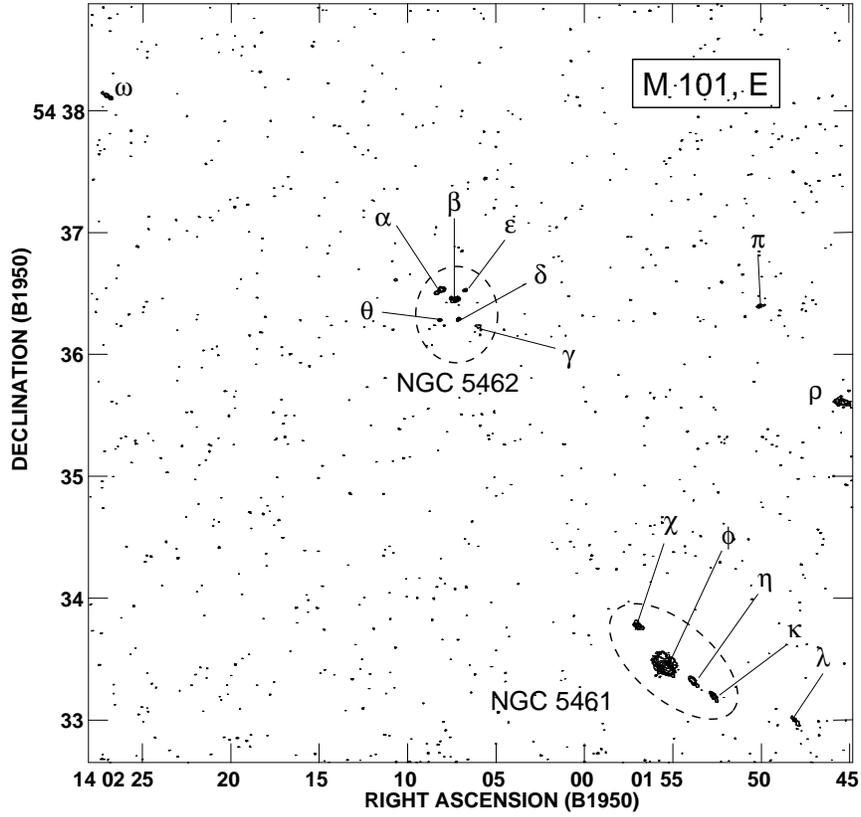}
\caption{
A contour map of part of the field of view for the eastern
region of NGC~5457 (M~101) at 20~cm, labeled as in Table~5.  Clearly
visible are the (circled) emission regions associated with NGC~5461 and
NGC~5462.  The contour levels are at $0.043$, $0.064$, $0.086$, $0.13$,
$0.21$, $0.43$, $0.64$ and $0.85$ mJy~beam$^{-1}$ and the rms noise is
$0.015$ mJy~beam$^{-1}$.
\label{fig9}}
\end{figure}






\clearpage

\begin{thebibliography}{}
\bibitem[Allen et al. 1976]{all76}
Allen, R. J., Goss, W. M., Ekers, R. D., \& de Bruyn, A. G.  1976, A\&A, 48,
253
\bibitem[Aretxaga et al. 1999]{are99}
Aretxaga, I., Benetti, S., Terlevich, R. J., Fabian, A. C., Cappellaro, E.,
Turatto, M., \& Della Valle, M. 1999, MNRAS, 309, 343
\bibitem[Baars et al. 1977]{baa77}
Baars, J. W. M., Genzel, R., Pauliny-Toth, I. I. K., \& Witzel, A. 1977, A\&A,
61, 99
\bibitem[Becker et al. 1995]{bec95}
Becker, R. H., White, R. L., \& Helfand, D. J. 1995, ApJ, 450, 559
\bibitem[Boffi \& Branch 1995]{bof95}
Boffi, F. R. \& Branch, D. 1995, PASP, 107, 347
\bibitem[Branch \& Cowan 1985]{bra85}
Branch, D., \& Cowan, J. J. 1985, ApJ, 297, L33
\bibitem[Branch et al. 1991]{bra91}
Branch, D., Nomoto, K., \& Filippenko, A. V. 1991, Comments Astrophys.,
15, 221
\bibitem[Branch et al. 1995]{bra95}
Branch, D., Livio, M., Yungelson, L. R., Boffi, F. R., \& Baron, E. 1995,
PASP, 107, 1019
\bibitem[Brown \& Marscher 1978]{bro78}
Brown, R. L., \& Marscher, A. P. 1978, ApJ, 220, 467
\bibitem[Cappellaro et al. 1995]{cap95}
Cappellaro, E., Danziger, I. J. \& Turatto, M. 1995, MNRAS, 277, 106
\bibitem[Chevalier 1982a]{che82a}
Chevalier, R. A. 1982a, ApJ, 258, 790
\bibitem[Chevalier 1982b]{che82b}
Chevalier, R. A. 1982b, ApJ, 259, 302
\bibitem[Chevalier 1984]{che84}
Chevalier, R. A. 1984, ApJ, 285, L63
\bibitem[Chevalier 1998]{che98}
Chevalier, R. A. 1998, AJ, 499, 810
\bibitem[Chu et al. 1999]{chu99}
Chu, Y-H., Caulet, A., Montes, M. J., Panagia, N., Van Dky, S. D., 
Weiler, K. W., 1999, ApJ, 512, L51
\bibitem[Clark \& Stephenson 1982]{cla82}
Clark, D. H., \& Stephenson, F., R. 1982, in Supernovae: A Survey of
Current Research, eds. M. J. Rees \& R. J. Stoneham (D. Reidel
Publishing Company), p. 355
\bibitem[Cowan \& Branch 1982]{cow82}
Cowan, J. J., \& Branch, D. 1982, ApJ, 258, 31
\bibitem[Cowan \& Branch 1985]{cow85}
Cowan, J. J., \& Branch, D. 1985, ApJ, 293, 400
\bibitem[Cowan, et al. 1988]{cow88}
Cowan, J. J., Henry, R. B. C., \& Branch. D. 1988, ApJ, 329, 116
\bibitem[Cowan et al. 1991]{cow91}
Cowan, J. J., Goss, W. M., \& Sramek, R. A. 1991, ApJ, 379, L49
\bibitem[Cowan et al. 1994]{cow94}
Cowan, J. J., Roberts, D. A., \& Branch, D. 1994, ApJ, 434, 128
\bibitem[Cowsik \& Sarkar 1984]{cow84}
Cowsik, R., \& Sarkar, S. 1984, MNRAS, 207, 745
\bibitem[Crane 1977]{cra77}
Crane, P. C. 1977, Thesis, Massachusetts Institute of Technology
\bibitem[Crane et al. 1992]{cra92}
Crane, P. C., Dickel, J. R. \& Cowan, J. J. 1992, ApJ, 390, L9
\bibitem[Donnelly et al. 1987]{don87}
Donnelly, R. H., Partridge, R. B. \& Windhorst, R. A. 1987, ApJ, 321, 94
\bibitem[Eck et al. 1995]{eck95}
Eck, C. R., Cowan, J. J.,  Roberts, D.,  Boffi, F. R., \& Branch, D.
1995, ApJ, 451, L53
\bibitem[Eck et al. 1996]{eck96}
Eck, C. R.,  Cowan, J. J.,  Boffi, F. R., \& Branch, D.
1996, ApJ,  472, L25
\bibitem[Eck et al. 1998]{eck98}
Eck, C. R., Roberts, D. , Cowan, J. J., \& Branch, D. 1998, ApJ, in press
\bibitem[Fesen et al. 1989]{fes89}
Fesen, R. A., Hamilton, A. J. S. \& Saken, J. M. 1989, ApJ, 341, L55
\bibitem[Fesen \& Becker 1990]{fes90}
Fesen, R. A. \& Becker, R. H. 1990, ApJ, 351, 437
\bibitem[Fesen 1993]{fes93a}
Fesen, R. A. 1993, ApJ, 413, L109
\bibitem[Fesen \& Matonick 1993]{fes93}
Fesen, R. A. \& Matonick, D. M. 1993, ApJ, 407, 110
\bibitem[Fesen et al. 1995]{fes95}
Fesen, R. A., Hurford, A. P. \& Matonick, D. M. 1995, AJ, 109, 2608
\bibitem[Fesen 1997]{fes97}
Fesen, R. A. 1997, BAAS, 29, 1268
\bibitem[Fesen 1998]{fes98}
Fesen, R. A. 1998, AJ, 115, 1107
\bibitem[Fesen et al. 1999]{fes99}
Fesen, R. A., Gerardy, G. L., Filippenko, A. V., Matheson, T., Chevalier, R. A.,
Kirshner, R. P., Schmidt, B. P., Challis, P., Fransson, C., Leibundgut, B., \&
Van Dyk, S. D., 1999, AJ, 117, 725
\bibitem[Filippenko et al. 1995]{fil95}
Filippenko, A. V., Barth, A. J., Bower, G. C., Ho, L. C., Stringfellow, G. S.,
Goodrich, R. W., \& Porter, A. C. 1995, AJ, 110, 2261
\bibitem[Goodrich et al. 1989]{goo89}
Goodrich, R. W., Stringfellow, G. S., Penrod, G. D., \& Filippenko, A. V.
1989, ApJ, 342, 908
\bibitem[Gottesman et al. 1972]{got72}
Gottesman, S. T., Broderick, J. J., Brown, R. L., Balick, B., \& Palmer, P.
1972, ApJ, 174, 383
\bibitem[Green 1984]{gre84}
Green, D. A. 1984, MNRAS, 209, 449
\bibitem[Gull 1973]{gul73}
Gull, S. F. 1973, MNRAS, 161, 47
\bibitem[Hartmann et al. 1986]{har86}
Hartmann, L. W., Geller, M. J., \& Huchra, J. P. 1986, AJ, 92(6), 1278
\bibitem[Hodge 1974]{hod74}
Hodge, P. W. 1974, ApJS, 27, 114
\bibitem[Hodge et al. 1990]{hod90}
Hodge, P. W., Gurwell, M., Goldader, J. D., \& Kennicutt, R. C., Jr. 1990,
ApJS, 73, 661
\bibitem[Hughes et al. 1998]{hug98}
Hughes, S. M. G., Mingsheng, H., Hoessel, J., Freedman, W. L., Kennicutt, Jr.,
R. C., Mould, J. R., Saha, A., Stetson, P. B., Madore, B. F., Silbermann,
N. A., Harding, P., Ferrarese, L., Ford, H., Gibson, B. K., Graham, J. A.,
Hill, R., Huchra, J., Illingworth, G. D., Phelps, R. \& Sakai, S.
1998, in press
\bibitem[Hummell et al. 1987]{hum87}
Hummell, E., van der Hulst, J. M., Keel, W. C., \& Kennicutt, R. C., Jr.
1987, A\&AS, 70, 517
\bibitem[Hyman et al. 1995]{hym95}
Hyman, S. D., Van Dyk, S. D., Weiler, K. W., \& Sramek, R. A. 1995,
ApJ, 443, L77
\bibitem[Israel et al. 1975]{isr75}
Israel, F. P., Goss, W. M., \& Allen, R. J. 1975, A\&A, 40, 421
\bibitem[Kelson et al. 1996]{kel96}
Kelson, D. D., Illingworth G. D., Freedman, W. F., Graham, J. A.,
Hill, R., Madore, B. F., Saha, A., Stetson, P. B., Kennicutt, Jr., R. C.,
Mould, J. R., Hughes, S. M., Ferrarese, L., Phelps, R., Turner, A.,
Cook, K. H., Ford, H., Hoessel, J. G. \& Huchra, J. 1996, ApJ,
463, 26
\bibitem[Lacey et al. 1999]{lac99}
Lacey, C. K., Weiler, K. W., Van Dyk, S. D., \& Sramek, R. A. 1999, BAAS, 
31, 976 
\bibitem[Leibundgut et al. 1991]{lei91}
Leibundgut, B., Kirshner, R. P., Pinto, P. A., Rupen, M. P., Smith, R. C.,
Gunn, J. E. \& Schneider, D. P. 1991, ApJ, 372, 531
\bibitem[Liller 1990]{lil90}
Liller, W. 1990, IBVS No. 3497
\bibitem[Long et al. 1989]{lon89}
Long, K. S., Blair, W. P. \& Krzeminski, W. 1989, ApJ, 340, L25
\bibitem[Long et al. 1992]{lon92}
Long, K. S., Winkler, P. F., \& Blair, W. P.  1992, ApJ, 395, 632
\bibitem[Lundqvist \& Fransson]{lun88}
Lundqvist, P. \& Fransson, C. 1988, A\&A, 192, 221
\bibitem[Madore \& Freedman 1991]{mad91}
Madore, B. F. \& Freedman, W. L. 1991, PASP, 103, 933
\bibitem[Matonick \& Fesen]{mat97}
Matonick, D. M. \& Fesen, R. A. 1997, ApJS, 112, 49
\bibitem[Montes et al. 1997]{mon97}
Montes, M. J., Van Dyk, S. D., Weiler, K. W., Sramek, R. A., \& Panagia, N.
1997, ApJ, 482, L61
\bibitem[Montes et al. 1998]{mon98}
Montes, M. J., Van Dyk, S. D., Weiler, K. W., Sramek, R. A., Panagia, N.,
1998, ApJ, 506, 874
\bibitem[Montes et al. 2000]{mon00}
Montes, M. J., Weiler, K. W., Van Dyk, S. D., Panagia, N., Lacey, C. K., 
Sramek, R. A., \& Park, R. 2000, ApJ, 532, ,1124
\bibitem[M\"{u}rset et al. 1991]{mur91}
M\"{u}rset, U., Nussbaumer, H., Schmidt, H. M., \& Vogel, M. 1991, A\&A,
248, 458
\bibitem[Nomoto et al. 1996]{nom96}
Nomoto, K., Iwamoto, K., Suzuki, T., Pols, O. R., Yamaoka, H., Hashimoto, M.,
H\"oflich, P., \& Van Den Heuvel, E. P. J. 1996, in Compact Stars in Binaries,
eds. J. van Paradijs et al., IAU (Netherlands), p. 119
\bibitem[Panagia et al. 1986]{pan86}
Panagia, N., Sramek, R. A., \& Weiler, K. W. 1986, ApJ, 300, L55
\bibitem[Romanishin 1997]{rom97}
Romanishin, W. 1997, private communication
\bibitem[Rupen et al. 1987]{rup87}
Rupen, M. P., van Gorkom, J. H., Knapp, G. R., Gunn, J. E., \& Schneider,
D. P. 1987, AJ, 94, 61
\bibitem[Ryder et al. 1993]{ryd93}
Ryder, S., Stavely-Smith, L., Dopita, M., Petre, R., Colbert, E.,
Malin, D., \& Schlegel, E. 1993, ApJ, 416, 167
\bibitem[Saha et al. 1994]{sah94}
Saha, A., Labhardt, L., Schwengeler, H., Macchetto, F. D.,
Panagia, N., Sandage, A., \& Tammann, G. A. 1994, ApJ, 425, 14
\bibitem[Saha et al. 1995]{sah95}
Saha, A., Sandage, A., Labhardt, L., Schwengler, H. \& Machetto, F. D.
1995, ApJ, 438, 8
\bibitem[Saha et al. 1997]{sah97}
Saha, A., Sandage, A., Labhardt, L., Tammann, G. A., Macchetto, F. D. \&
Panagia, N. 1997, ApJ, 486, 1
\bibitem[Sandage \& Tammann 1974]{san74}
Sandage, A. \& Tammann, G. A., 1974, ApJ, 194, 223
\bibitem[Sandage \& Tammann 1982]{san82}
Sandage, A. \& Tammann, G. A., 1982, ApJ, 256, 339
\bibitem[Schlegel 1990]{sch90}
Schlegel, E. M. 1990, MNRAS, 244, 269
\bibitem[Schlegel et al. 1999]{sch99}
Schlegel, E. M., Ryder, S., Staveley-Smith, L., Petre, R., Colbert, E.,
Dopita, M., Campbell-Wilson, D. 1999, AJ, 118, 2689
\bibitem[Seaquist \& Taylor 1990]{sea90}
Seaquist, E. R., \& Taylor, A. R. 1990, ApJ, 349, 313
\bibitem[Silbermann et al. 1996]{sil96}
Silbermann, N. A., Harding, P., Madore, B. F.,
Kennicutt, Jr., R. C., Saha, A., Stetson, P. B., Freedman, W. L., Mould,
J. R., Graham, J. A., Hill, R. J., Turner, A., Bresolin, F., Ferrarese, L.,
Ford, H., Hoessel, J. G., Han, M., Huchra, J., Hughes, S. M. G.,
Illingworth, G. D., Phelps, R. \& Sakai, S. 1996, ApJ, 470, 1
\bibitem[Sivan et al. 1990]{siv90}
Sivan, J. -P., Petit, H., Compte, G., \& Maucherat, A. J. 1990, A\&A,
237, 23
\bibitem[Sonneborn et al. 1998]{son98}
Sonneborn, G., Pun, C. S. J., Kimble, R. A., Gull, T. R., Lundqvist, P.,
McCray, R., Plait, P., Boggess, A., Bowers, C. W., Danks, A. C.,
Grady, J., Heap, S. R., Kraemer, S., Lindler, D., Loiacono, J., Maran,
S. P., Moos, H. W.,\& Woodgate, B. E., ApJ, 492, L139
\bibitem[Sramek et al. 1984]{sra84}
Sramek, R. A., Panagia, N., \& Weiler, K. W. 1984, ApJ, 285, L59
\bibitem[Sramek et al. 1993]{sra93}
Sramek, R. A., Weiler, K. W., Van Dyk, S., \& Panagia, N. 1993, in
Sub-Arcsecond Radio Astronomy, eds. R. J. Davis \& R. S. Booth (Cambridge,
Cambridge Un. Press), p. 32
\bibitem[Stockdale et al. 2001a]{sto01a}
Stockdale, C. J.,   Goss, W. M.,  Cowan, J. J., \&  Sramek, R. A.
2001, ApJ, submitted
\bibitem[Stockdale et al. 2001b]{sto01b}
Stockdale, C. J.,  Rupen, M. P.,  Cowan, J. J., Chu, Y.-H. \&  Jones, S.
2001, AJ, in press
\bibitem[Tully 1988]{tul88}
Tully, R. B. 1988, The Catalog of Nearby Galaxies (Cambridge, Cambridge
University Press)
\bibitem[Turner \& Ho 1994]{tur94}
Turner, J. L., \& Ho, P. T. 1994, ApJ, 421, 122
\bibitem[Van Dyk et al. 1992]{van92}
Van Dyk, S. D., Weiler, K. W., Sramek, R. A. \& Panagia, N. 1992, ApJ, 396, 195
\bibitem[Van Dyk et al. 1993a]{van93a}
Van Dyk, S. D., Sramek, R. A., Weiler, K. W., \& Panagia, N. 1993a,
ApJ, 409, 162
\bibitem[Van Dyk et al. 1993b]{van93b}
Van Dyk, S. D., Weiler, K. W., Sramek, R. A., \& Panagia, N., 1993b,
ApJ, 419, L69
\bibitem[Van Dyk et al. 1994]{van94}
Van Dyk, S. D., Weiler, K. W., Sramek, R. A., Rupen, M., \& Panagia, N., 1994,
ApJ, 432, L115
\bibitem[Van Dyk 1997]{van97}
Van Dyk, S. 1997, private communication
\bibitem[Van Dyk et al. 1998]{van98}
Van Dyk, S. D., Montes, M. J., Weiler, K. W., Sramek, R. A., \& Panagia, N., 1998,
AJ, 115, 1103
\bibitem[Weiler et al. 1989]{wei89}
Weiler, K. W., Panagia, N., Sramek, R. A., van der Hulst, J. M.,
Roberts, M. S. \& Nguyen, L. 1989, ApJ, 336, 421
\bibitem[Weiler et al. 1990]{wei90}
Weiler, K. W., Panagia, N. \& Sramek, R. A. 1990, ApJ, 364, 611
\bibitem[Weiler \& Sramek]{wei88}
Weiler, K. W., \& Sramek, R. A., 1988, ARAA, 26, 295
\bibitem[Weiler et al. 1986]{wei86}
Weiler, K. W., Sramek, R. A., Panagia, N., van der Hulst, J. M.,
\& Salvati, M. 1986, ApJ, 301, 790
\bibitem[Weiler et al. 1998]{wei98}
Weiler, K. W., van Dyk, S. D., Montes, M., Panagia, N., \& Sramek, R. A. 
1998, ApJ, 500, 51
\bibitem[Weiler et al. 1991]{wei91}
Weiler, K. W., van Dyk, S. D., Panagia, N., Sramek, R. A., \&
Discenna, J. L. 1991, ApJ, 380, 161
\bibitem[Weiler et al. 1992a]{wei92a}
Weiler, K. W., Van Dyk, S. D., Panagia, N., \& Sramek, R. A. 1992a, ApJ, 398,
248
\bibitem[Weiler et al. 1992b]{wei92b}
Weiler, K. W., Van Dyk, S. D., Pringle, J. E., \& Panagia, N. 1992b,
ApJ, 399, 672
\bibitem[Willis et al. 1976]{wil76}
Willis, A. G., Oosterbaan, C. E., de Ruiter, H. R. 1976, A\&ASup, 25, 453
\bibitem[Zhang et al. 1993]{zha93}
Zhang, X., Wright, M., Alexander, P. 1993, ApJ, 418, 100
\end{thebibliography}
\end{document}



\def\fsec {{\rlap.}^{\rm s}}
\def\fasec {{\rlap.}^{\prime\prime}} \hskip0.05em
\def\asec {^{\prime\prime}}
\def\fmin {{\rlap.}^{\rm m}}
\def\famin {{\rlap.}^{\prime}} \hskip0.05em
\def\deg {^\circ}
\def\amin {^{\prime}}
\def\lldots {\multicolumn{1}{l}{$\cdots$}}

\footnotesize

\begin{table}

\noindent\begin{tabular}{clcclccr@{$\;$}lr@{$\;$}lr@{$\;$}ll} 

\multicolumn{14}{c}{Table 3.  Summary of data on intermediate-age SNe and 
younger SNe in our survey.} \\

\multicolumn{7}{c}{ } & \multicolumn{6}{c}{Flux Density (mJy)$^c$} 
& \multicolumn{1}{c}{ } \\ 
\cline{9-14}

	& SN position$^{a}$ & & & & & & & & & & & & \\ 

SN	& \multicolumn{1}{c}{$\alpha_{1950}$} & SN & Parent 
& \multicolumn{1}{c}{$\alpha_{1950}^{a}$} & 
Galaxy & Age & \multicolumn{2}{c}{20 cm} &  
\multicolumn{2}{c}{$\:$6 cm} & \multicolumn{2}{c}{Other} &  \\ 

Name    & \multicolumn{1}{c}{$\delta_{1950}$} & Type  & Galaxy
& \multicolumn{1}{c}{$\delta_{1950}$} & 
Type$^{a}$ & (yr)$^{b}$ & 
& & & & 
\multicolumn{2}{c}{$\lambda$} & \multicolumn{1}{c}{Notes} \\ 
\tableline\tableline

SN 1885A & $\;\;\;00^{\rm h}39^{\rm m}58\fsec 6$ & I pec & NGC 0224 & 
$\;\;\;00^{\rm h}40^{\rm m}00\fsec 1$ & SA(s)b & 105 & & \lldots & 
 & \lldots & $<$ & 0.015 & 1, 2 \\

   & $+40\deg 59\amin 38\asec$ & & (M31) & $+40\deg 59\amin 42\fasec 8$ 
                                               & & & & & & & & & \\

SN 1895B & $\;\;\;13^{\rm h}37^{\rm m}06\fsec 6$ & Ia & NGC 5253 & 
$\;\;\;13^{\rm h}37^{\rm m}05\fsec 1$ & Im pec & 85.8 & 
$<$ & 0.9 & & \lldots & & \lldots & 3, 4 \\ 

   & $-31\deg 23\amin 01\asec$ & & & $-31\deg 23\amin 13\fasec 2$ 
              & & 89.5 & $<$ & 0.15 & & \lldots & & \lldots & \\

   & & & & & & & & & & & & & \\

SN 1909A & $\;\;\;14^{\rm h}00^{\rm m}17\fasec 2$ & II(-P) & NGC 5457 & 
$\;\;\;14^{\rm h}01^{\rm m}26\fsec 3$ & SAB(rs)cd & 79.7 & & \lldots 
 & $<$ & 0.034 & & \lldots & 3 \\

   & $+54\deg 42\famin 22\asec$ & & (M101) & $+54\deg 35\amin 17\fasec 9$ & 
 					 & & & & & & & & \\

SN 1921B & $\;\;\;10^{\rm h}15^{\rm m}19\fsec 6$ & II & NGC 3184 & 
$\;\;\;10^{\rm h}15^{\rm m}17\fsec 2$ & SAB(rs)cd & 62.7 & $<$ & 0.51 
 & & \lldots & & \lldots & 3, 4 \\

   & $+41\deg 37\amin 20\asec$ & & & $+41\deg 40\amin 29\fasec 0$ & & &
                                                             & & & & & & \\

SN 1921C & $\;\;\;10^{\rm h}15^{\rm m}23\fsec 7$ & I & NGC 3184 & 
$\;\;\;10^{\rm h}15^{\rm m}17\fsec 2$ & SAB(rs)cd & 62.0 & $<$ & 0.51 & 
 & \lldots & & \lldots & 4, 5 \\

   & $+41\deg 36\amin 04\asec$ & & & $+41\deg 40\amin 29\fasec 0$ & & &
                                                             & & & & & & \\

SN 1923A & $\;\;\;13^{\rm h}34^{\rm m}20\fsec 0$ & II-P & NGC 5236 &
$\;\;\;13^{\rm h}34^{\rm m}11\fsec 6$ & SAB(s)c & 60.6 & $<$ & 0.57 
 & & \lldots & & \lldots & \\

   & $-29\deg 35\amin 48\fasec 0$ & & (M83) & $-29\deg 36\amin 42\fasec 2$ 
           & & 60.8 & & \lldots & & 0.19 $\pm$ 0.05 & & \lldots & \\

   & & & & & & 67.4 & & \lldots & & 0.093 $\pm$ 0.03 & & \lldots & \\

   & & & & & & 68.6 & & 0.30 $\pm$ 0.05 & & \lldots & & \lldots & \\

   & & & & & & & & & & & & & \\

SN 1937C & $\;\;\;13^{\rm h}03^{\rm m}33\fsec 1$ & Ia & IC 4182 & 
$\;\;\;13^{\rm h}03^{\rm m}30\fsec 2$ & SA(s)m & 48.7 & $<$ & 0.10 & 
 & \lldots & & \lldots & \\

   & $+37\deg 53\amin 10\asec$ & & & $+37\deg 52\amin 20\fasec 0$ & & &
                                                                & & & & & & \\

SN 1937F & $\;\;\;10^{\rm h}15^{\rm m}17\fsec 2$ & II(-P) & NGC 3184 & 
$\;\;\;10^{\rm h}15^{\rm m}17\fsec 2$ & SAB(rs)cd & 46.0 & $<$ & 0.51 & 
& \lldots & & \lldots & 4 \\

   & $+41\deg 37\amin 31\fasec$ & & & $+41\deg 40\amin 29\fasec 0$ & & & 
                                                             & & & & & & \\

SN 1945B & $\;\;\;13^{\rm h}34^{\rm m}02\fsec 8$ & ? & NGC 5236 &
$\;\;\;13^{\rm h}34^{\rm m}11\fsec 6$ & SAB(s)c & 
   38.4 & $<$ & 0.57 & & \lldots & & \lldots & \\

   & $-29\deg 39\amin 43\asec$ & & (M83) & $-29\deg 36\amin 42\fasec 2$ 
           & & 38.7 & & \lldots & $<$ & 0.16 & & \lldots & \\

   & & & & & & 45.2 & & \lldots & $<$ & 0.084 & & \lldots & \\

   & & & & & & 46.5 & $<$ & 0.15 & & \lldots & & \lldots & \\

   & & & & & & & & & & & & & \\

SN 1950B & $\;\;\;13^{\rm h}34^{\rm m}03\fsec 7$ & ? & NGC 5236 &
$\;\;\;13^{\rm h}34^{\rm m}11\fsec 6$ & SAB(s)c & 
   33.8 & & 0.7 $\pm$ 0.08 & & \lldots & & \lldots & \\

   & $-29\deg 36\amin 39\fasec 0$ & & (M83) & $-29\deg 36\amin 42\fasec 2$ 
           & & 34.0 & & \lldots & & 0.36 $\pm$ 0.04 & & \lldots & \\

   & & & & & & 40.6 & & \lldots & & 0.37 $\pm$ 0.03 & & \lldots & \\

   & & & & & & 41.8 & & 0.72 $\pm$ 0.04 & & \lldots & & \lldots & \\

   & & & & & & & & & & & & & \\

SN 1951H & $\;\;\;14^{\rm h}02^{\rm m}07\fsec 2$ & II & NGC 5457 & 
$\;\;\;14^{\rm h}01^{\rm m}26\fsec 3$ & SAB(rs)cd & 42.1 & $<$ 
 & $0.044$ & & \lldots & & \lldots & 3 \\

   & $+54\deg 36\amin 21\asec$ & & (M101) & $+54\deg 35\amin 17\fasec 9$ & 
            	 			& & & & & & & & \\

SN 1954A & $\;\;\;12^{\rm h}13^{\rm m}15\fsec 2$ & Ib & NGC 4214 &
$\;\;\;12^{\rm h}13^{\rm m}08\fsec 0$ & IAB(s)m & 32.0 & $<$ & 0.068 & 
& \lldots & & \lldots & \\

   & $+36\deg 32\amin 54\asec$ & & & $+36\deg 36\amin 22\fasec 0$ & &
                                                              & & & & & & & \\

SN 1954J & $\;\;\;07^{\rm h}32^{\rm m}08\fsec 6$ & V & NGC 2403 & 
$\;\;\;07^{\rm h}32^{\rm m}05\fsec 5$ & SAB(s)cd & 30.8 & $<$ & 0.35 & 
 & \lldots & & \lldots & 3, 6 \\

   & $+65\deg 44\famin 22\asec$ & & & $+65\deg 42\amin 40\fasec 0$ & 
 					& & & & & & & &  \\

SN 1957D & $\;\;\;13^{\rm h}34^{\rm m}14\fsec 4$ & ? & NGC 5236 &
$\;\;\;13^{\rm h}34^{\rm m}11\fsec 6$ & SAB(s)c & 
   26.0 & & 2.7 $\pm$ 0.1 & & \lldots & & \lldots & \\

   & $-29\deg 34\amin 24\fasec 3$ & & (M83) & $-29\deg 36\amin 42\fasec 2$ 
           & & 26.2 & & \lldots & & 2.1 $\pm$ 0.06 & & \lldots & \\

   & & & & & & 32.8 & & \lldots & & 1.4 $\pm$ 0.04 & & \lldots & \\

   & & & & & & 34.0 & & 1.2 $\pm$ 0.07 & & \lldots & & \lldots & \\

   & & & & & & & & & & & & & \\ 










SN 1959D & $\;\;\;22^{\rm h}34^{\rm m}44\fsec 4$ & II-L & NGC 7331 & 
$\;\;\;22^{\rm h}34^{\rm m}46\fsec 7$ & SA(s)b & 26.1 & $<$ & 0.058 
 & & \lldots & & \lldots & 3 \\

   & $+34\deg 09\amin 33\fasec 1$ & & & $+34\deg 09\amin 20\fasec 9$ 
           & & 26.8 & $<$ & 0.079 & & \lldots & & \lldots & \\

   & & & & & & 27.1 & & \lldots & $<$ & 0.042 & & \lldots & \\ 

   & & & & & & & & & & & & & \\

SN 1961V & $\;\;\;02^{\rm h}40^{\rm m}29\fsec 7$ & II pec (V) & NGC 1058 &
$\;\;\;02^{\rm h}40^{\rm m}23\fsec 2$ & SA(rs)c & 22.9 & 
 & 0.27 $\pm$ 0.02 & & & & & 3, 7 \\

   & $+37\deg 08\amin 01\fasec 8$ & & & $+37\deg 07\amin 45\fasec 0$ 
           & & 24.7 & & \lldots & & 0.13 $\pm$ 0.01 & & \lldots &  \\

   & & & & & & & & & & & & & \\

SN 1966B & $\;\;\;12^{\rm h}45^{\rm m}12\fsec 1$ & II-L & NGC 4688 &
$\;\;\;12^{\rm h}45^{\rm m}14\fsec 0$ & SB(s)cd & 21.5
                                 & $<$ & 0.16 & & \lldots & & \lldots & \\

   & $+04\deg 35\amin 58\asec$     & & & $+04\deg 36\amin 27\fasec 0$ 
                                        & & & & & & & & &  \\

SN 1968L & $\;\;\;13^{\rm h}34^{\rm m}11\fsec 3$ & II-P & NGC 5236 &
$\;\;\;13^{\rm h}34^{\rm m}11\fsec 6$ & SAB(s)c & 
   15.4 & $<$ & 0.57 & & \lldots & & \lldots & \\

   & $-29\deg 36\amin 42\fasec 6$ & & (M83) & $-29\deg 36\amin 42\fasec 2$ 
           & & 15.7 & & \lldots & $<$ & 0.16 & & \lldots & \\
 
   & & & & & & 22.3 & & \lldots & $<$ & 0.084 & & \lldots & \\
   
   & & & & & & 23.5 & $<$ & 0.15 & & \lldots & & \lldots & \\

   & & & & & & & & & & & & & \\

SN 1969L & $\;\;\;02^{\rm h}40^{\rm m}38\fsec 6$ & II-P & NGC 1058 & 
$\;\;\;02^{\rm h}40^{\rm m}23\fsec 2$ & SA(rs)c & 14.9 & $<$ & 0.09 
				& & \lldots & & \lldots & 3, 10 \\

   & $+37\deg 05\amin 58\asec$ & & & $+37\deg 07\amin 45\fasec 0$ 
           & & 16.7 & & \lldots & $<$ & 0.042 & & & \\

SN 1970G & $\;\;\;14^{\rm h}01^{\rm m}14\fsec 4$ & II-L & NGC 5457 &
$\;\;\;14^{\rm h}01^{\rm m}26\fsec 3$ & SAB(rs)cd & 19.8 
   & & 0.21 $\pm$ 0.02 & & \lldots & & \lldots & 3 \\

   & $+54\deg 28\amin 55\fasec 8$ & & (M101) & $+54\deg 35\amin 17\fasec 9$ 
           & & 20.3 & & \lldots & & \lldots & & 0.076 $\pm$ 0.01 & 2 \\

   & & & & & & & & & & & & & \\

SN 1972E & $\;\;\;13^{\rm h}37^{\rm m}02\fsec 4$ & Ia & NGC 5253 & 
$\;\;\;13^{\rm h}37^{\rm m}05\fsec 1$ & Im pec & 9.0 & $<$ & 0.9 & 
 & \lldots & & \lldots & 3, 4 \\

   & $-31\deg 25\amin 04\asec$ & & & $-31\deg 23\amin 13\fasec 2$
      & & 13.6 & $<$ & 0.15 & & \lldots & & \lldots & \\

   & & & & & & & & & & & & & \\

SN 1973R & $\;\;\;11^{\rm h}17^{\rm m}35\fsec 0$ & II-P & NGC 3627 & 
$\;\;\;11^{\rm h}17^{\rm m}38\fsec 5$ & SAB(s)b & 13.6 
                             & $<$ & 0.28 & & \lldots & & \lldots & 3 \\

   & $+13\deg 16\amin 13\asec$ & & (M66) & $+13\deg 15\amin 55\fasec 6$
                                        & & & & & & & & & \\

SN 1980D & $\;\;\;11^{\rm h}32^{\rm m}13\fsec 4$ & II-P & NGC 3733 & 
$\;\;\;11^{\rm h}32^{\rm m}17\fsec 1$ & SAB(s)cd & 7.4 & $<$ 
 & 0.094 & & \lldots & & \lldots & \\

   & $+55\deg 09\amin 30\fasec 0$ & & & $+55\deg 07\amin 43\fasec 0$ 
					& & & & & & & & & \\

SN 1983K & $\;\;\;12^{\rm h}46^{\rm m}36\fsec 4$ & IIB-P & NGC 4699 &
$\;\;\;12^{\rm h}46^{\rm m}26\fsec 5$  & SAB(rs)b & 4.1 & & \lldots & 
                                   $<$ & 0.052 & & \lldots & 8 \\

   & $-08\deg 21\amin 21\fasec 0$ & & & $-08\deg 23\amin 34\asec$
                                         & & & & & & & & & \\

SN 1983N & $\;\;\;13^{\rm h}34^{\rm m}02\fsec 0$ & Ib & NGC 5236 & 
$\;\;\;13^{\rm h}34^{\rm m}11\fsec 6$ & SAB(s)c & 
   0.47 & & 4.4 $\pm$ 0.2 & & \lldots & & \lldots & \\

   & $-29\deg 38\amin 45\fasec 2$ & & (M83) & $-29\deg 34\amin 42\fasec 4$
           & & 0.70 & & \lldots & & 0.52 $\pm$ 0.05 & & \lldots & \\

   & & & & & & 7.3 & & \lldots & $<$ & 0.084 & & \lldots & \\

   & & & & & & 8.5 & $<$ & 0.15 & & \lldots & & \lldots & \\

   & & & & & & & & & & & & & \\

SN 1984E & $\;\;\;10^{\rm h}11^{\rm m}35\fsec 4$  & II-L & NGC 3169 &
$\;\;\;10^{\rm h}11^{\rm m}39\fsec 4$  & SA(s)a pec & 11.3 
 & $<$ & 0.073 & & \lldots & & \lldots & \\

   & $+03\deg 43\amin 07\fasec 7$ & & & $+03\deg 42\amin 52\fasec 0$
                                         & & & & & & & & & \\

SN 1986E & $\;\;\;12^{\rm h}19^{\rm m}09\fsec 0$  & II-L & NGC 4302 &
$\;\;\;12^{\rm h}19^{\rm m}10\fsec 1$  & Sc & 9.3 & $<$ & 0.17 & 
 & \lldots & & \lldots & \\

   & $+14\deg 54\amin 33\fasec 0$ & & & $+14\deg 52\amin 30\fasec 3$
          & & & & \lldots & $<$ & 0.038 & & \lldots & \\ 

   & & & & & & & & & & & & & \\

SN 1986G & $\;\;\;13^{\rm h}22^{\rm m}40\fsec 4$  & Ia & NGC 5128 &
$\;\;\;13^{\rm h}22^{\rm m}31\fsec 6$  & S0 pec & -0.02 
                                  & & \lldots & $<$ & 0.96 & & \lldots & \\

   & $-42\deg 46\amin 18\fasec 0$    & & & $-42\deg 45\amin 32\fasec 9$
          & & & & \lldots & & \lldots & $<$ & 0.69 & 9 \\

   & & & & & & & & & & & & & \\

SN 1989B & $\;\;\;11^{\rm h}17^{\rm m}37\fsec 4$ & Ia & NGC 3627 &
$\;\;\;11^{\rm h}17^{\rm m}38\fsec 5$ & SAB(s)b & -0.008
                             & & \lldots & $<$ & 0.11 & & \lldots & \\

   & $+13\deg 16\amin 45\fasec 0$ & & (M66) & $+13\deg 15\amin 55\fasec 6$
          & & & & \lldots & & \lldots & $<$ & 0.089 & 2 \\
\tableline

\end{tabular}

a. Supernovae and galaxy positions and types taken from NED. Positions 
have been rounded to the nearest $0\fsec 1$ and $0\fasec 1$ 
when applicable. \newline
b. Age of the supernova past maximum brightness or past date of discovery
(if the date of maximum is not known) at the time of observation.  \newline
A negative age stands for a pre-maximum observation. \newline
c. Flux densities are $3\sigma$ upper limits. \newline
\end{table}
\pagebreak
\footnotesize
Notes \newline
1) This upper limit provided by Crane et al.\ (1992). \newline
2) Observation was made at $3.6$ cm. \newline
3) Brown \& Marscher (1978) report radio observations of 
these SNe resulting in upper limits to detection. \newline 
4) The upper limits from observations of these SNe are quoted 
from Cowan \& Branch (1982) and Branch \& Cowan (1985) since a map 
\newline was unavailable for independent analysis. \newline
5) There is some dicrepancy with the NED reported position for SN 1921C 
and the astrometric position reported in Branch \& Cowan (1985). \newline
6) There is not much evidence that the explosion 
associated with SN 1954J was an actual SN.  \newline
7) It is not certain that SN 1961V was actually an SN explosion (Fillipenko
et al. 1995). \newline
8) NGC 4699 is also home to SN 1948A, but its position lies outside of 
the field of view for our map. \newline
9) Observation was made at 2 cm. \newline
10) The 20 cm upper limit for SN 1969L is quoted from Branch \& Cowan (1985) 
since a map was unavailable for independent analysis. \newline



\def\fsec {{\rlap.}^{\rm s}}
\def\fasec {{\rlap.}^{\prime\prime}} \hskip0.05em
\def\asec {^{\prime\prime}}
\def\fmin {{\rlap.}^{\rm m}}
\def\famin {{\rlap.}^{\prime}} \hskip0.05em
\def\deg {^\circ}
\def\amin {^{\prime}}
\def\dots {\multicolumn{1}{c}{$\cdots$}}

\footnotesize

\begin{table}
\noindent\begin{tabular}{lcr@{:}llrrlcc} 

\multicolumn{10}{c}{Table 5.  Summary of observational data on non-SNe 
sources.} \\

\multicolumn{1}{c}{Galaxy} & \multicolumn{1}{c}{Galaxy} & 
\multicolumn{1}{c}{ } & 
\multicolumn{2}{c}{Position of Peak$^{a}$} & 
\multicolumn{3}{c}{Peak Flux (mJy beam$^{-1}$)} & & \\
\cline{6-8}

\multicolumn{1}{c}{Name} & \multicolumn{1}{c}{Type} & 
\multicolumn{1}{c}{ } &
\multicolumn{1}{c}{$\alpha_{1950}$} & 
\multicolumn{1}{c}{$\delta_{1950}$} & 
\multicolumn{1}{c}{20 cm} & \multicolumn{1}{c}{6 cm} & 
\multicolumn{1}{c}{3.6 cm} & Classification? & Notes \\ \tableline\tableline \\

NGC 1058 & Sc & $\beta$ & $02^{\rm h}40^{\rm m}31\fsec 4$ & 
   $+37\deg 09\amin 03\fasec 0$ & \dots & $0.10 \pm 0.01$ 
 & \dots & & 1 \\ 

 & & $\alpha$ & $02^{\rm h}40^{\rm m}45\fsec 9$ & 
   $+37\deg 07\amin 21\fasec 7$ & \dots & $0.25 \pm 0.01$ 
 & \dots & \\ \\

NGC 2403 & Scd & $\phi$ & $07^{\rm h}31^{\rm m}29\fsec 4$ & 
   $+65\deg 46\amin 10\fasec 9$ & $1.2 \pm 0.1$ & \dots & \dots & &  \\

 & & $\eta$ & $07^{\rm h}31^{\rm m}30\fsec 7$ &
   $+65\deg 43\amin 46\fasec 1$ & $1.3 \pm 0.1$ & \dots & \dots & H II & 2 \\

 & & $\psi$ & $07^{\rm h}31^{\rm m}32\fsec 1$ & 
   $+65\deg 47\amin 58\fasec 8$ & $2.3 \pm 0.1$ & \dots & \dots & &  \\

 & & $\chi$ & $07^{\rm h}31^{\rm m}39\fsec 9$ & 
   $+65\deg 40\amin 30\fasec 6$ & $0.75 \pm 0.1$ & \dots & \dots & H II &  \\ 

 & & $\omega$ & $07^{\rm h}31^{\rm m}52\fsec 7$ & 
   $+65\deg 43\amin 32\fasec 3$ & $0.87 \pm 0.1$ & \dots & \dots & H II & 2 \\

 & & $\epsilon$ & $07^{\rm h}31^{\rm m}56\fsec 4$ & 
   $+65\deg 43\amin 41\fasec 9$ & $1.4 \pm 0.1$ & \dots & \dots & H II & 2 \\

 & & $\mu$ & $07^{\rm h}31^{\rm m}57\fsec 0$ &
   $+65\deg 43\amin 19\fasec 5$ & $0.62 \pm 0.1$ & \dots & \dots & SNR? \\

 & & $\theta$ & $07^{\rm h}32^{\rm m}00\fsec 1$ &
   $+65\deg 43\amin 32\fasec 6$ & $0.72 \pm 0.1$ & \dots & \dots & H II & 2 \\

 & & $\delta$ & $07^{\rm h}32^{\rm m}03\fsec 4$ &
   $+65\deg 43\amin 28\fasec 1$ & $1.6 \pm 0.1$ & \dots & \dots & H II & 2 \\

 & & $\pi$ & $07^{\rm h}32^{\rm m}07\fsec 1$ & 
   $+65\deg 42\amin 24\fasec 5$ & $0.61 \pm 0.1$ & \dots & \dots & &  \\

 & & $\gamma$ & $07^{\rm h}32^{\rm m}08\fsec 7$ & 
   $+65\deg 44\amin 05\fasec 7$ & $1.5 \pm 0.1$ & \dots & \dots & H II & 2 \\

 & & $\beta$ & $07^{\rm h}32^{\rm m}18\fsec 1$ & 
  $+65\deg 43\amin 21\fasec 5$ & $4.4 \pm 0.1$ & \dots & \dots & H II & 2 \\

 & & $\rho$ & $07^{\rm h}32^{\rm m}28\fsec 0$ & 
   $+65\deg 45\amin 13\fasec 2$ & $0.78 \pm 0.1$ & \dots & \dots & H II &  \\

 & & $\xi$ & $07^{\rm h}32^{\rm m}29\fsec 6$ & 
   $+65\deg 40\amin 32\fasec 7$ & $0.92 \pm 0.1$ & \dots & \dots & H II &  \\

 & & $\alpha$ & $07^{\rm h}32^{\rm m}36\fsec 9$ & 
   $+65\deg 43\amin 22\fasec 1$ & $1.9 \pm 0.1$ & \dots & \dots & & \\ \\

NGC 3169 & Sa pec & C & $10^{\rm h}11^{\rm m}39\fsec 4$ & 
   $+03\deg 42\amin 52\fasec 5$ & $3.4 \pm 0.02$ & \dots & \dots & 
       GC & 3 \\

 & & $\alpha$ & $10^{\rm h}11^{\rm m}41\fsec 7$ & 
   $+03\deg 41\amin 37\fasec 2$ & $0.67 \pm 0.02$ & \dots & \dots 
 & & \\ \\

NGC 3627 & SAB(s)b & $\gamma$ & $11^{\rm h}17^{\rm m}36\fsec 5$ &
   $+13\deg 14\amin 03\fasec 3$ & $2.1 \pm 0.07$ & \dots 
 & \dots & & 4 \\

 & & $\pi$ & $11^{\rm h}17^{\rm m}36\fsec 7$ &
   $+13\deg 17\amin 01\fasec 8$ & \dots & \dots & $0.17 \pm 0.03$ & 
       & 5 \\ 

 & & $\epsilon$ & $11^{\rm h}17^{\rm m}36\fsec 9$ &
   $+13\deg 16\amin 52\fasec 0$ & \dots & $0.47 \pm 0.04$ 
 & $0.25 \pm 0.03$ & & 6 \\

 & & $\alpha$ & $11^{\rm h}17^{\rm m}38\fsec 5$ &
   $+13\deg 15\amin 55\fasec 7$ & $8.1 \pm 0.07$ & $3.2 \pm 0.04$
 & \dots & GC & 7 \\

 & & $\beta$ & $11^{\rm h}17^{\rm m}40\fsec 1$ & 
   $+13\deg 15\amin 09\fasec 8$ & $1.3 \pm 0.07$ & \dots 
 & \dots & H II & 8 \\ \\

NGC 4214 & Im & $\delta$ & $12^{\rm h}13^{\rm m}08\fsec 3$  & 
   $+36\deg 35\amin 39\fasec 0$ & $0.13 \pm 0.03$ & \dots 
 & \dots & & 9 \\

 & & $\gamma$ & $12^{\rm h}13^{\rm m}08\fsec 4$ & 
   $+36\deg 36\amin 16\fasec 8$ & $0.32 \pm 0.03$ & \dots 
 & \dots & H II \\

 & & $\epsilon$ & $12^{\rm h}13^{\rm m}08\fsec 6$  & 
   $+36\deg 36\amin 13\fasec 4$ & $0.59 \pm 0.02$ & \dots 
 & \dots \\

 & & $\eta$ & $12^{\rm h}13^{\rm m}08\fsec 9$ & 
   $+36\deg 36\amin 12\fasec 5$ & $0.09 \pm 0.02$ & 
   \dots & \dots & H II & 10 \\ 

 & & $\pi$ & $12^{\rm h}13^{\rm m}09\fsec 3$ & 
   $+36\deg 35\amin 20\fasec 7$ & $0.34 \pm 0.03$ & \dots 
 & \dots \\

 & & $\beta$ & $12^{\rm h}13^{\rm m}10\fsec 1$ & 
   $+36\deg 35\amin 49\fasec 6$ & $0.80 \pm 0.03$ & \dots 
 & \dots & H II \\

 & & $\alpha$ & $12^{\rm h}13^{\rm m}10\fsec 3$ & 
   $+36\deg 35\amin 44\fasec 0$ & $0.82 \pm 0.02$ & \dots 
 & \dots & H II \\

 & & $\rho$ & $12^{\rm h}13^{\rm m}11\fsec 3$ & 
   $+36\deg 35\amin 55\fasec 0$ & $0.20 \pm 0.03$ & \dots 
 & \dots \\






 & & $\omega$ & $12^{\rm h}13^{\rm m}19\fsec 1$ & 
   $+36\deg 35\amin 27\fasec 2$ & $0.32 \pm 0.03$ & \dots & 
        \dots & & \\

 & & $\theta$ & $12^{\rm h}13^{\rm m}20\fsec 5$ & 
   $+36\deg 35\amin 24\fasec 8$ & $0.16 \pm 0.03$ & \dots & 
       \dots & & \\

 & & $\xi$ & $12^{\rm h}13^{\rm m}26\fsec 5$ & 
   $+36\deg 35\amin 09\fasec 5$ & $0.43 \pm 0.02$ & \dots &
       \dots & & \\ \\
 
NGC 4302 & Sc & $\delta$ & $12^{\rm h}19^{\rm m}01\fsec 3$ &
   $+14\deg 53\amin 51\fasec 3$ & \dots & $0.44 \pm 0.02$ 
 & \dots & & 11 \\ 

 & & $\beta$ & $12^{\rm h}19^{\rm m}09\fsec 8$ & 
   $+14\deg 54\amin 51\fasec 2$ & $1.1 \pm 0.07$ & $2.0 \pm 0.03$
 & \dots & & 12 \\

 & & $\gamma$ & $12^{\rm h}19^{\rm m}14\fsec 0$ &
   $+14\deg 53\amin 35\fasec 2$ & $0.84 \pm 0.07$ & 
 $2.2 \pm 0.02$ & \dots &  & 13 \\ \\

NGC 4688 & SBcd & $\alpha$ & $12^{\rm h}45^{\rm m}11\fsec 7$ & 
   $+04\deg 35\amin 27\fasec 5$ & $0.96 \pm 0.07$ & \dots & \dots & 
 & 14 \\ \\

NGC 5128 & S0 pec & C & $13^{\rm h}22^{\rm m}31\fsec 7$ 
 & $-42\deg 44\amin 59\fasec 5$ & \dots & \multicolumn{1}{c}{5660} 
 & $^{2cm}$131 & GC & 15 \\ \\







NGC 5457, S & Scd & $\psi$ & $14^{\rm h}00^{\rm m}42\fsec 9$ & 
   $+54\deg 30\amin 39\fasec 9$ & \dots & \dots & $0.07 \pm 0.01$ & H II 
 & 22 \\

 & & $\theta$ & $14^{\rm h}00^{\rm m}44\fsec 0$ & 
   $+54\deg 30\amin 33\fasec 6$ & \dots & \dots & $0.06 \pm 0.01$ & H II 
 & 22 \\ 

 & & $\rho$ & $14^{\rm h}01^{\rm m}00\fsec 2$ & 
  $+54\deg 29\amin 18\fasec 6$ & $0.11 \pm 0.02$ & \dots & \dots & H II 
 & 16, 22 \\

 & & $\pi$ & $14^{\rm h}01^{\rm m}08\fsec 7$ & 
   $+54\deg 28\amin 44\fasec 9$ & $0.18 \pm 0.02$ & \dots 
 & $0.06 \pm 0.01$ & \\

 & & $\epsilon$ & $14^{\rm h}01^{\rm m}10\fsec 5$ & 
   $+54\deg 30\amin 42\fasec 0$ & $1.1 \pm 0.02$ & \dots 
 & $0.39 \pm 0.01$ & H II & 22 \\

 & & $\mu$ & $14^{\rm h}01^{\rm m}11\fsec 0$ & 
   $+54\deg 27\amin 13\fasec 2$ & $0.10 \pm 0.02$ & \dots & \dots & & 16 \\

 & & $\eta$ & $14^{\rm h}01^{\rm m}12\fsec 2$ &
   $+54\deg 30\amin 25\fasec 8$ & $0.14 \pm 0.02$ & \dots 
 & \dots & & 16 \\

 & & $\alpha$ & $14^{\rm h}01^{\rm m}14\fsec 7$ & 
   $+54\deg 28\amin 52\fasec 1$ & $0.69 \pm 0.02$ & \dots 
 & $0.72 \pm 0.01$ & H II & 17 \\

 & & $\gamma$ & $14^{\rm h}01^{\rm m}19\fsec 4$ & 
   $+54\deg 29\amin 12\fasec 9$ & $0.51 \pm 0.02$ & \dots 
 & $0.19 \pm 0.01$ & \\
 
 & & $\omega$ & $14^{\rm h}01^{\rm m}20\fsec 3$ & 
   $+54\deg 29\amin 03\fasec 9$ & $0.10 \pm 0.01$ & \dots 
 & $0.09 \pm 0.01$ & H II & 18 \\

 & & $\delta$ & $14^{\rm h}01^{\rm m}23\fsec 5$ & 
   $+54\deg 30\amin 04\fasec 2$ & $0.14 \pm 0.02$ & \dots 
 & $0.11 \pm 0.01$ & H II & 19, 22 \\ \\

NGC 5457, E & Scd & $\rho$ & $14^{\rm h}01^{\rm m}45\fsec 6$ & 
   $+54\deg 35\amin 36\fasec 3$ & $0.18 \pm 0.02$ & \dots & \dots
 & \\

 & & $\lambda$ & $14^{\rm h}01^{\rm m}48\fsec 1$ & 
   $+54\deg 32\amin 59\fasec 9$ & $0.094 \pm 0.02$ & \dots 
 & \dots & \\
 
 & & $\pi$ & $14^{\rm h}01^{\rm m}50\fsec 1$ & 
   $+54\deg 36\amin 23\fasec 9$ & $0.12 \pm 0.02$ & \dots 
 & \dots & H II & 23 \\

 & & $\kappa$ & $14^{\rm h}01^{\rm m}52\fsec 7$ & 
   $+54\deg 33\amin 11\fasec 9$ & $0.12 \pm 0.01$ & \dots 
 & \dots & H II & 24 \\

 & & $\eta$ & $14^{\rm h}01^{\rm m}53\fsec 8$ & 
   $+54\deg 33\amin 19\fasec 2$ & $0.17 \pm 0.02$ & \dots 
 & \dots & H II & 23, 24 \\

 & & $\phi$ & $14^{\rm h}01^{\rm m}55\fsec 4$ & 
   $+54\deg 33\amin 26\fasec 6$ & $0.78 \pm 0.01$ & \dots 
 & \dots & H II & 23, 24  \\

 & & $\chi$ & $14^{\rm h}01^{\rm m}57\fsec 0$ & 
   $+54\deg 33\amin 46\fasec 5$ & $0.10 \pm 0.01$ & \dots 
 & \dots & H II & 24 \\

 & & $\gamma$ & $14^{\rm h}02^{\rm m}06\fsec 1$ & 
   $+54\deg 36\amin 14\fasec 0$ & $0.057 \pm 0.02$ & \dots 
 & \dots & H II & 23, 24 \\

 & & $\epsilon$ & $14^{\rm h}02^{\rm m}06\fsec 8$ & 
   $+54\deg 36\amin 31\fasec 6$ & $0.11 \pm 0.02$ & \dots 
 & \dots & H II & 23, 24 \\

 & & $\delta$ & $14^{\rm h}02^{\rm m}07\fsec 1$ & 
   $+54\deg 36\amin 17\fasec 2$ & $0.13 \pm 0.02$ & \dots 
 & \dots & H II & 24 \\

 & & $\beta$ & $14^{\rm h}02^{\rm m}07\fsec 2$ & 
   $+54\deg 36\amin 27\fasec 3$ & $0.20 \pm 0.02$ & \dots 
 & \dots & H II & 24 \\

 & & $\alpha$ & $14^{\rm h}02^{\rm m}08\fsec 1$ & 
   $+54\deg 36\amin 32\fasec 0$ & $0.18 \pm 0.01$ & \dots 
 & \dots & H II & 23, 24 \\

 & & $\theta$ & $14^{\rm h}02^{\rm m}08\fsec 2$ & 
   $+54\deg 36\amin 17\fasec 2$ & $0.10 \pm 0.02$ & \dots 
 & \dots & \\

 & & $\omega$ & $14^{\rm h}02^{\rm m}26\fsec 9$ & 
   $+54\deg 38\amin 06\fasec 3$ & $0.15 \pm 0.01$ & \dots 
 & \dots & \\ \\

IC 4182 & Sm & $\alpha$ & $13^{\rm h}03^{\rm m}12\fsec 9$ & 
   $+37\deg 51\amin 59\fasec 5$ & $0.48 \pm 0.02$ & \dots 
 & \dots & & 20 \\

 & & $\beta$ & $13^{\rm h}03^{\rm m}12\fsec 4$ & 
   $+37\deg 51\amin 53\fasec 9$ & $1.4 \pm 0.03$ & \dots 
 & \dots & & 21 \\
\\ \tableline\tableline

\end{tabular}
\newline
a. The sources are designated according to the label on each map. GC 
is the designation for Galactic Center. 
\end{table}

\pagebreak
\footnotesize

Notes \newline
1) Comparison of the 20 cm map with an H$\alpha$ image of the galaxy 
reveals no source coincident with source $\beta$. \newline
2) The position of each of these sources is within $1\asec - 5\asec$ 
of a corresponding source found by Turner \& Ho (1994). \newline
3) This emission region is associated with the central regions of 
NGC 3169. \newline
4) This source is relatively far from the center of NGC 3627.  \newline
5) This source is only visible at 3.6 cm due to the reduced noise
at that wavelength.  It has a position within $4\asec$ of an H II 
region listed \newline by Hodge (1974) (see Appendix A). \newline
6) This source is not visible at 20 cm because of reduced noise level at 6 and 
3.6 cm. The reduced flux at 3.6 cm relative to 6 cm 
is \newline indicative of non-thermal emission, but this source is  
coincident with an H II region from Hodge (1974) (see Appendix A). \newline 
7) The non-thermal spectral index for this source and the coincidence
with the central regions of NGC 3627 indicate this region may \newline be 
associated with the dynamic center for the galaxy. \newline
8) The amorphous shape of this extended emission region and its 
positional coincidence over several Hodge (1974) H II regions 
indicate \newline this source is probably one or more H II regions. \newline
9) Source $\delta$ has a peak flux of only 4.5$\sigma$. \newline
10) Source $\eta$ seems to be a complex composed of several distinguishable 
smaller sources. The flux and position listed is for the strongest 
\newline source. \newline 
11) This source was not visible at 20 cm due to increased noise at 
that wavelength. \newline
12) The spectral index for this source, $+0.6$, 
is very similar to those of background radio galaxies and/or 
quasars found in \newline Donnelly, Partridge, \& Windhorst (1987). \newline
13) The spectral index for this source, $+0.4$, 
is also similar to background radio galaxies and/or quasars
in Donnelly et al. (1987). \newline
14) This source is also extended and has an amorphous shape.  Perhaps 
it is actually several sources.  \newline
15) The source detected here is resolved at both wavelengths. The integrated
fluxes are 6060 and 380 mJy at 6 and 2 cm respectively. 
\newline The resulting scaled array spectral index is -2.5 \newline
16) These sources did not have a signal above $4\sigma$ at 3.6 cm.\newline 
17) This source is coincident with the position of NGC 5455, an HII 
region. While $\alpha$ is $3\fasec 6$ outside of the uncertainty 
($\pm 10\asec$) of a \newline source (No. 416) 
reported to be associated with 
NGC 5455 in Hodge et al. (1990), our source has an extended structure 
\newline (see Figure 8) which makes the identification likely.  \newline
18) The flat spectral index ($\alpha^{20}_{3.6} = -0.052$) indicates 
probable identification as an HII region.\newline
19) The flat spectral index for this source ($\alpha^{20}_{3.6} = 
+0.026$) indicates thermal emission characteristic of HII regions. \newline
20) This source was observed by the FIRST survey at the VLA at 21.4 cm 
and is somewhat amorphous.  The flux reported by FIRST \newline is 
about $2.5$ times larger than our peak flux. \newline 
21) Also observed in the FIRST survey, this source is extended and has 
a jet-like shape that points towards/away from source $\alpha$. 
The \newline peak flux observed by us is very similar to that reported 
by FIRST. \newline
22) This source has a position that is within the uncertainties for an H II 
region found by Hodge et al. (1990).  We list our source and \newline 
the corresponding 
H II region identification in Hodge et al. (1990): $\psi$ - No. 141, 
$\rho$ - No. 217, $\theta$ - No. 158, $\epsilon$ - No. 365, 
$\delta$ - No. 592. \newline
23) These sources have positions within the uncertainties for H II regions
found by Hodge et al. (1990). \newline
24) Superposition of the radio map of NGC 5461 \& 5462 with H$\alpha$ maps 
in Israel, Goss, \& Allen (1975) has these sources lying over H$\alpha$ 
emission regions. \newline